\documentclass{aa}

\usepackage{natbib}
\usepackage{amsmath}    
\usepackage{subfigure}
\usepackage{graphicx}
\usepackage{float}
\usepackage{natbib}
\usepackage{dcolumn}
\usepackage{bm}
\usepackage{ulem}
\usepackage{color}
\usepackage{txfonts}
\usepackage{rotating}
\usepackage{booktabs}
\usepackage{array}
\usepackage{tabularx}
\usepackage{threeparttable}
\usepackage{diagbox}
\usepackage{multirow}
\usepackage{makecell}
\usepackage{longtable}
\usepackage{mhchem}
\usepackage{afterpage}
\usepackage{hyperref}

\begin{document}

\title{The mixing of internal gravity waves and lithium production in intermediate-mass AGB stars}

\author{Zhijun Wang\inst{1}
\and    Guoliang L\"{u}\inst{1*}
\and    Chunhua Zhu\inst{1*}
\and    Sufen Guo\inst{1}
\and    Helei Liu\inst{1}
\and    Xizhen Lu\inst{1}
}
 \institute{School of Physical Science and Technology, Xinjiang University, Urumqi, 830046, China\\
              \email{guolianglv@sina.com};{chunhuazhu@sina.cn}}

  \abstract
   {Intermediate-mass asymptotic giant branch (AGB) stars may have significant influence in the evolution of lithium (Li) in the Galaxy. During the AGB phase, stars eject surface material into the interstellar matter (ISM) by stellar winds, and the Li content in their surfaces and winds decisively influences the AGB stars contribution of Li to the ISM. Turbulent convection within stars, driven by internal gravity waves (IGW) excited by convective motions, can transmit energy outward and induce mixing in non-convective regions, profoundly affecting the chemical composition of stellar surfaces and winds.}
   {We investigate the effects of IGW on Li production of the AGB stars. We will demonstrate the validity of extra-mixing triggered by IGW in the radiative zone between convective thermal pulse and the convective envelope of Intermediate-mass AGB stars, and investigate its impact on Li production in these stars. In this paper we will use this model combined with initial mass functions to derive the total Li production from intermediate-mass AGB stars.}
   {We use the Modules for Experiments in Stellar Astrophysics to construct stellar models from the zero-age main sequence to the end of AGB. This simulation incorporated IGW-induced mixing coefficients and element diffusion effects. Grids are established to calculate the production of Li for AGB stars with different masses and metallicities. Subsequently, employing the method of population synthesis, we use linear interpolation in grids to estimate the contribution rate of Li production to the Galactic from a sample size of $10^7$ stars.}
   {Simulated results demonstrate that IGW triggered during each Helium-shell flash induces extra-mixing into non-convective regions. In our model, IGW induces an extra-mixing that transports material from the radiative zone between the convective thermal pulse and the convective envelope to the convective envelope, allowing \ce{^7Be} originally located below the convective envelope to be transported into the convective envelope where \ce{^7Be} decays into \ce{^7Li}. The positive effects of IGW mixing on Li yield diminish with increasing initial mass. For models with the same initial mass, the positive impact of IGW mixing increases with metallicity. In our calculations, most of AGB stars which initial masses between 3.5$M_\odot$ and 7.5$M_\odot$ can produce a positive Li yields. By the method of synthesis population, we estimate the total Li yields produced by AGB stars with IGW mixing is about 15$M_\odot$, which is twice as much as that without IGW mixing. The contribution of these models to total galactic Li is around 10\%. It means that AGB star may be a non-negligible source for Li production.}
   {Through this extra-mixing mechanism induced by IGW, AGB stars can achieve a maximum A(Li) exceeding 5 and intermediate-mass AGB stars significantly contribute to Li in the Galactic ISM. These findings underscore the crucial role of IGW in stellar evolution, particularly in enhancing Li production.}

   \keywords{Stars: evolution ; stars: AGB and post-AGB ; stars:abundances; ISM:abundances
   }

   \maketitle
\section{Introduction}
Lithium (Li) is one of the three elements formed during the Big Bang. As the age of the universe increases, the abundance of Li in the interstellar matter (ISM) has risen from the primordial Li abundance formed during the Big Bang, A(Li)=2.72 \citep{1998PhRvA..58..883Z,2014JCAP...10..050C} to the current meteorites abundance A(Li)=3.26 \citep{2009ARA&A..47..481A,2009LanB...4B..712L}. Pranztos (\citeyear{2012A&A...542A..67P}) considered half of the Galactic Li comes from stars, this includes low-mass red giants (initial mass between $1 \sim 2$$M_\odot$), asymptotic giant branch (AGB) stars (initial mass between $2 \sim 6$$M_\odot$) and novae. Otherwise, AGB stars with initial mass over 6$M_\odot$ also exhibits higher Li production \citep{2010MNRAS.402L..72V,2012MSAIS..22..247L}. This indicates that AGB stars with masses between $2 \sim 8$$M_\odot$ are all potential contributors to significant Li enrichment in the Galaxy, especially the intermediate-mass AGB stars capable of undergoing the Hot Bottom Burning (HBB) process, with initial masses between 3$M_\odot$ (depending on metallicities) to 8$M_\odot$ \citep{2001ApJ...559..909T,2016ApJ...825...26K}.
Current research results indicate that novae explosions contribute the most to the production of Li in the Galactic. Estimates from various studies have suggested a lower bound 10\% \citep{2017PASP..129g4201R,2024ApJ...962..191S} to higher bound 70\% \citep{2024ApJ...971....4G}. Li-rich Giant (which A(Li) > 1.5) \citep{1967ApJ...147..624I} could also be a potential source of Li. Extensive observational evidence confirms that low-mass red giants during Red Clump after the Helium flash(He-flash) can also enter a Li-rich or even super Li-rich phase \citep{2018NatAs...2..790Y,2019MNRAS.484.2000D,2019ApJ...878L..21S,2020JApA...41...49K,2021ApJ...913L...4S,2021MNRAS.505.5340M}. However, they have low mass-loss rate, and their Li-rich phase only lasts for a short duration.
Stellar models of intermediate initial mass that experience AGB are considered to be potential sources of Li \citep{1992ApJ...392L..71S,2009A&A...499..835V}. In the stellar interior, the production of Li primarily comes from the decay of \ce{^7Be} and the production of \ce{^7Be} depends on the $^{3}\text{He}(\alpha{},{} \gamma)^{7}\text{Be}$ reaction which occurs at the temperature over $4 \times 10^{7}$ K, $^{7}\text{Be}(p,\alpha{})^{4}\text{He}$ occurs at the temperature over $8 \times 10^{7}$ K, and \ce{^7Be} decays into \ce{^7Li} through nuclear reactions $^{7}\text{Be}(p,\alpha{})^{4}\text{He}$ occurs at the temperature over $8 \times 10^{7}$ K, and \ce{^7Be} decays into \ce{^7Li} through nuclear reactions $^{7}\text{Be}(e, \nu)^{7}\text{Li}$ with a half-life of 53 days at low temperature \citep{1955ApJ...122..271F,1955ApJ...121..144C,2013ApJ...764..118S}. Otherwise, \ce{^7Li} is destroyed when the temperature higher than $2.5 \times 10^{6}$K. It means that \ce{^7Li} survives hardly in the stellar interior where \ce{^7Be} is produced. Therefore, Cameron \& Fowler proposed that \ce{^7Be} produced in the stellar interior must be transported rapidly into cool envelope (where the temperature is lower than $2.5 \times 10^{6}$ K) via convection to prevent the decay-generated \ce{^7Li} from being destroyed, which is called as CF mechanism \citep{1971ApJ...164..111C}. Intermediate-mass stars during the thermal pulses asymptotic giant branch (TP-AGB) have temperatures at the bottom of their convective envelopes high enough to trigger nuclear reactions producing \ce{^7Be} and the CNO cycle, leading to the Hot bottom burning (HBB) process. These stars transport nuclear reaction products to the stellar surface via convection, resulting in Li-rich and O-rich (C/O < 1 in surface). The discovery of Li-rich AGB stars with S-process enrichment and O-rich compositions in the Magellanic Cloud (MC) has confirmed this theory \citep{1983ApJ...272...99W,1989ApJ...345L..75S,1990ApJ...361L..69S,1993ApJ...418..812P}.
In previous investigations, the Li production in AGB stars was found to be influenced by parameters such as convection, mixing and mass-loss rates, higher mass-loss rates generally lead to greater Li production \citep{2001ApJ...559..909T,2001A&A...374..646R}. Based on the IRAS two-colour diagram, \citet{2007A&A...462..711G,2013A&A...555L...3G} and \citet{2019A&A...623A.151P} detected the Li abundances of 30 O-rich AGB stars in the Milky Way. They found that 14 of them are Li-rich AGB stars, and the highest A(Li) among these stars reaches 4.3 dex. Spectral analysis indicates that all samples within the Milky Way exhibit a mass-loss rate of less than $10^{-6}~M_{\odot}$yr$^{-1}$\citep{2014A&A...564L...4Z,2017A&A...606A..20P,2019A&A...623A.151P}. The presence of Li-rich and O-rich confirm that these stars have undergone the HBB process. This observational result confirms the unreliability of simulation results under conditions of high mass-loss rates.
The treatment of convection has also been a critical factor influencing Li production. Bigger mixing length parameters lead to higher efficiency of convection \citep{1996ApJ...473..550C}. On the other hand, theoretically, Full Spectrum of Turbulence models \citep{1991ApJ...370..295C} with higher convection efficiency also exhibit higher A(Li) during the HBB process \citep{1999A&A...348..846M,2005A&A...431..279V}. Furthermore, some extra-mixing mechanisms have been shown to effectively enhance surface \ce{^7Li} abundance during the AGB, such as thermohaline mixing \citep{2010IAUS..268..405S,2021MNRAS.505.5340M}, proton ingestion events in low-mass stars \citep{2004ApJ...602..377I,2024Galax..12...66C} and rotationally-induced mixing \citep{1998A&A...335..959T}. Magnetic and internal gravitational waves (IGW) are also considered as possible sources of extra-mixing mechanism \citep{2010sf2a.conf..253L}
During TP-AGB stage stars have both the convective thermal pulse and the convective envelope. Through the convective envelope, they transport nuclear reaction products from the interior to the surface and eventually eject these products into the ISM. The convective thermal pulse primarily occurs within the Helium shell (He-shell) during Helium burning (He-burning). Stars in TP-AGB experience a cycle where the He-shell is repeatedly ignited until the outer envelope are completely ejected. With each pulse, the peak luminosity during He-shell burning increases.

The IGW triggered by turbulent convection within stars is considered one of the reasons for the onset of extra-mixing mechanism \citep{1994A&A...281..421M}. The ignition process of the helium shell in stars during AGB creates conditions for the excitation of IGW. This leads to the generation of an extra-mixing coefficient in the non-convective zones of stars, bringing \ce{^7Li} to the surface and ultimately enhancing Li production. To our knowledge, there is seldom any investigations about the effect of IGW on Li yields for intermediate-mass AGB stars.

In this paper, we demonstrate that the excitation of IGW during He-shell burning can induce mixing in the radiative zone between the convective thermal pulse to the convective envelope. We provide the production of Li and its contribution to the Galactic Li mass under reliable mass loss rate simulations. Section 2 introduces model parameters and demonstrate the validity of IGW-induced mixing. Section 3 demonstrates the impact of mixing mechanisms on Li production in AGB stars and compare our results with observations. In Section 4, we presents the conclusions of this paper.

\section{Model}
We use Modules for Experiments in Stellar Astro-physics (MESA,[rev22.11.1]; \citep{2011ApJS..192....3P,2013ApJS..208....4P,2015ApJS..220...15P,2018ApJS..234...34P,2019ApJS..243...10P,2021zndo...5798242P})to construct one-dimension intermediate-mass stellar models. MESA adopts the equation of state of Rogers \& Nayfonov (\citeyear{2002ApJ...576.1064R}) and Timmes \& Swesty (\citeyear{2000ApJS..126..501T}) and the opacity of Iglesias \& Rogers (\citeyear{1993ApJ...412..752I,1996ApJ...464..943I}) and Ferguson (\citeyear{2005ApJ...623..585F}).
\begin{figure}[htbp]
    \centering
    \begin{minipage}[b]{0.4\textwidth}
        \centering
        \includegraphics[width=\textwidth]{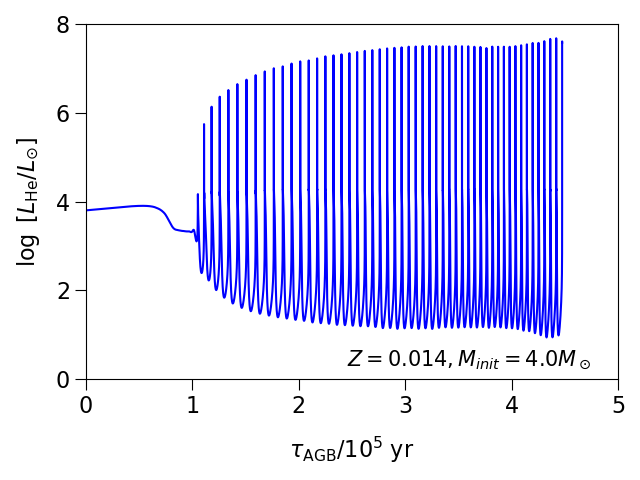}
        \label{fig:24}
    \end{minipage}
    \hfill
    \begin{minipage}[b]{0.42\textwidth}
        \centering
        \includegraphics[width=\textwidth]{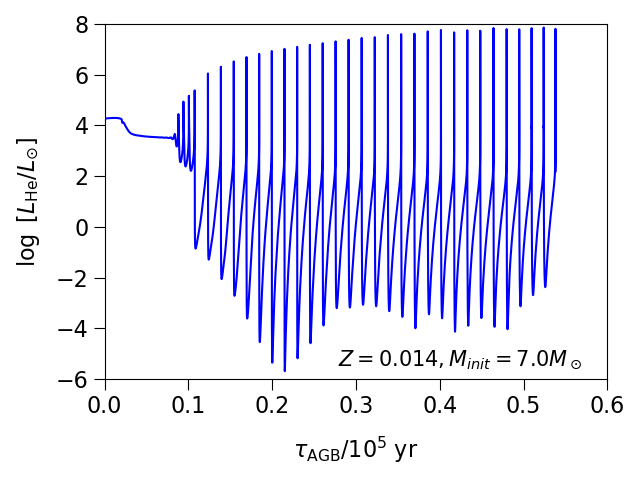}
        \label{fig:25}
    \end{minipage}
    \caption{Helium-luminosity over time during TP-AGB for the $Z$=0.014, 4$M_\odot$ and 7$M_\odot$ models. For the x-axis, $\mathrm{\tau}_{\text{AGB}}$ represents evolutionary time after the begining of AGB phase at which the core He-burning is just exhausted.}
    \label{fig:four}
\end{figure}
\subsection{Model Parameters}
Different metallicities are selected as $Z$=0.014, $Z$=0.004, $Z$=0.0014, $Z$=0.00014. Stellar initial mass in the models are 3.5$M_\odot$ to 7.5$M_\odot$ depending on different metallicity. The convective efficiency inside stars depends on parameter settings. In our model, the mixing length parameter is set as 1.9.
\subsubsection{Overshooting}
Exponential overshooting bottom of both the convective thermal pulse and the convective envelope also be considered. According to Herwig (\citeyear{2000A&A...360..952H}) and Freytag (\citeyear{1996A&A...313..497F}), the diffusion coefficient in the
overshoot region is:
\begin{equation}
D_{OV} = D_0 \exp\left(-\frac{2z}{H_v}\right), \quad H_v = f H_p.
\label{eq:6}
\end{equation}
Here $D_0$ is the diffusion coefficient at the boundary of convection zone, $z$ is the distance to the boundary of convection zone, $H_p$ is the pressure scale height. Following Herwig (\citeyear{1999A&A...349L...5H}), the overshooting parameter $f$ is set to 0.016.
\subsubsection{Unclear network}
For nuclear network, we use AGB.net which includes nuclear reactions from \ce{^1H} to \ce{^{22}Ne}. We adopt nuclear reaction rates compiled by JINA REACLIB (\citeyear{2010ApJS..189..240C}). $^{7}\text{Be}(e, \nu)^{7}\text{Li}$ reaction is adopt Simonucci (\citeyear{2013ApJ...764..118S}), as made available in machine-readable form by Vescovi (\citeyear{2019A&A...623A.126V}). This reaction rate incorporates a revision to the electron capture rate at temperatures below $10^7$K. Treatment of electron screening is based on Alastuey \& Jancovici (\citeyear{1978ApJ...226.1034A}) and Itoh (\citeyear{1979ApJ...234.1079I}).
\subsubsection{Mass loss prescription}
The mass-loss formula in \citet{1995A&A...297..727B} (B95) is adopted during AGB, the mass-loss rate is given by
\begin{equation}
\dot{M} = 4.83\cdot 10^{-9}M^{-2.1}L^{-2.7}\dot{M_R},
\label{eq:7}
\end{equation}
which $M_R$ is given by \citet{1975psae.book..229R}
\begin{equation}
\dot{M_R} = 4\cdot 10^{-13}\dot{\eta}\frac{LR}{M}.
\label{eq:8}
\end{equation}
Here, $\eta$ is a dimensionless free parameter scaling the mass-loss efficiency. We set $\eta$=0.01, this value has been adopted in multiple previous models \citep{2001A&A...374..646R,2000A&A...363..605V}.
\subsubsection{Element diffusion mechanism}
Element diffusion is also considered, this can partially explain the mechanism behind the production of Li-rich giants near the tip of the red giant branch (RGB). The enrichment of Li in these stars can be explained by considering the element diffusion mechanism \citep{2015ApJS..220...15P,2018ApJS..234...34P} within a certain constant diffusion coefficient. This further explains the production mechanism of super Li-rich giants \citep{2022A&A...668A.126G}. In stellar interiors, the primary factors influencing element diffusion in stars include the pressure gradient (gravity), temperature gradient, composition gradient, and radiation pressure, with gravitational sedimentation being the main driving force. MESA employs the method proposed by \citet{1994ApJ...421..828T}, which solves the entire system of equations defined in \citet{1969fecg.book.....B} within a matrix structure without imposing restrictions on the number of elements considered. By inputting the number density $n_s$, temperature $T$, gradients ${\mathrm{d} \ln n_s}/{\mathrm{d} r}$, ${\mathrm{d} \ln T}/{\mathrm{d} r}$, material mass (in atomic mass units $A_s$, mean charge $Z_s$ of the material in its average ionization state, and resistive coefficients $K_{st}$, $z_{st}'$, and $z_{st}''$ \citep{2015ApJS..220...15P}, MESA computes elemental diffusion in stellar interiors, incorporating specific diffusion coefficients derived from \citet{1986ApJS...61..177P} and \citet{2016PhRvE..93d3203S}.
\subsubsection{Initial composition}
The contribution of stars to the Galactic Li inventory depends on both Li production during stellar evolution and Li consumption from the ISM during star formation. The initial Li mass in stars is calculated by multiplying the stellar initial mass by the initial Li abundance, which varies with metallicity. In the present paper, according to the observations, the
initial A(Li) is set to 3.26 for stars with $Z$=0.014, 2.24 for stars with low-metallicities ($Z$=0.00014 or $Z$=0.0014) \citep{1997MNRAS.292L...1M,1996A&A...307..172S}, and 2.5 for
stars with $Z=0.004$ \citep{2000AJ....120.1841F,2001ApJ...559..909T}. All models begin their evolution from the zero-age main sequence.
\subsection{Extra-Mixing Excited by IGW}
During TP-AGB stage, the He-shell undergoes repeated burning and extinguishment cycles. The peak Helium-luminosity during each burning event gradually increases with each subsequent burning (He-shell flash). During He-shell flash, the energy produced by the flash powers a convective region (the convective thermal pulse) in the He-burning shell \citep{2014PASA...31...30K}. The Hydrogen-rich convective envelope remains in a convective state (the convective envelope). A radiative zone exists between the convective thermal pulse and the convective envelope throughout the period from the He-shell flash to the third dredge-up process.
IGW is considered to be triggered by turbulent convective motions within the interiors of stellars \citep{1981ApJ...245..286P,2013MNRAS.430.2363L}, the energy transmitted outward by IGW, after overcoming the gravitational potential of a certain region, can lead to that region being in a mixed state. The IGW model has successfully explained variations in Li abundance in solar and low-mass stars \citep{1991ApJ...377..268G,2005Sci...309.2189C}. IGW triggered during the first He-flash in red giants can induce mixing between Hydrogen-burning zone and the convective envelope, and bring \ce{^7Be} in the Hydrogen-burning zone into the convective envelope \citep{2020ApJ...901L..18S}. This can partially explain the mechanism behind the production of Li-rich giants near the tip of the red giant branch (RGB). The existence of super Li-rich giants can be further explained by additionally incorporating the the element diffusion primarily driven by gravitational settling \citep{2022A&A...668A.126G} and the neutrino magnetic moment-induced extra Cooling Effect \citep{2025PhRvD.111j3004L}.
\subsubsection{Evidence that IGW can trigger extra-mixing}
Here we consider that IGW triggered by He-shell burning during the TP-AGB can also induce mixing in the radiative zone between the convective thermal pulse and the convective envelope. In Fig.\ref{fig:four}, we present the plot showing the Helium-luminosity evolution over time for $Z$=0.014, 4$M_\odot$ and 7$M_\odot$ stars during TP-AGB. After several initial pulses, the maximum He-burning luminosity during each pulse exceed $10^{7} \textit{L}_\odot $. In the final pulses they may even reach $10^{8} \textit{L}_\odot$.
Press (\citeyear{1981ApJ...245..286P}) investigated the mixing induced by IGW in the stellar interior. In their model, a region with mixing height $H$ and volume $V$ must overcome gravitational potential energy, which requires energy provided by IGW of $H^2N^2\rho V$, where $\rho$ is density and $N$ is Brunt-V\"{a}is\"{a}l\"{a} frequency. To sustain the mixed state, this energy must be updated within the thermal diffusion timescale $H^2/K$. Dividing the two quantities yields the power required to maintain mixing is $N^2K\rho V$, which is independent of height $H$. For unit mass, this value becomes $N^2K$. Therefore, the power $L_{\text {mix }}$ required to sustain a constant mass mixing state should be equal to:
\begin{figure}[htbp]
\centering
\includegraphics[width=0.5\textwidth]{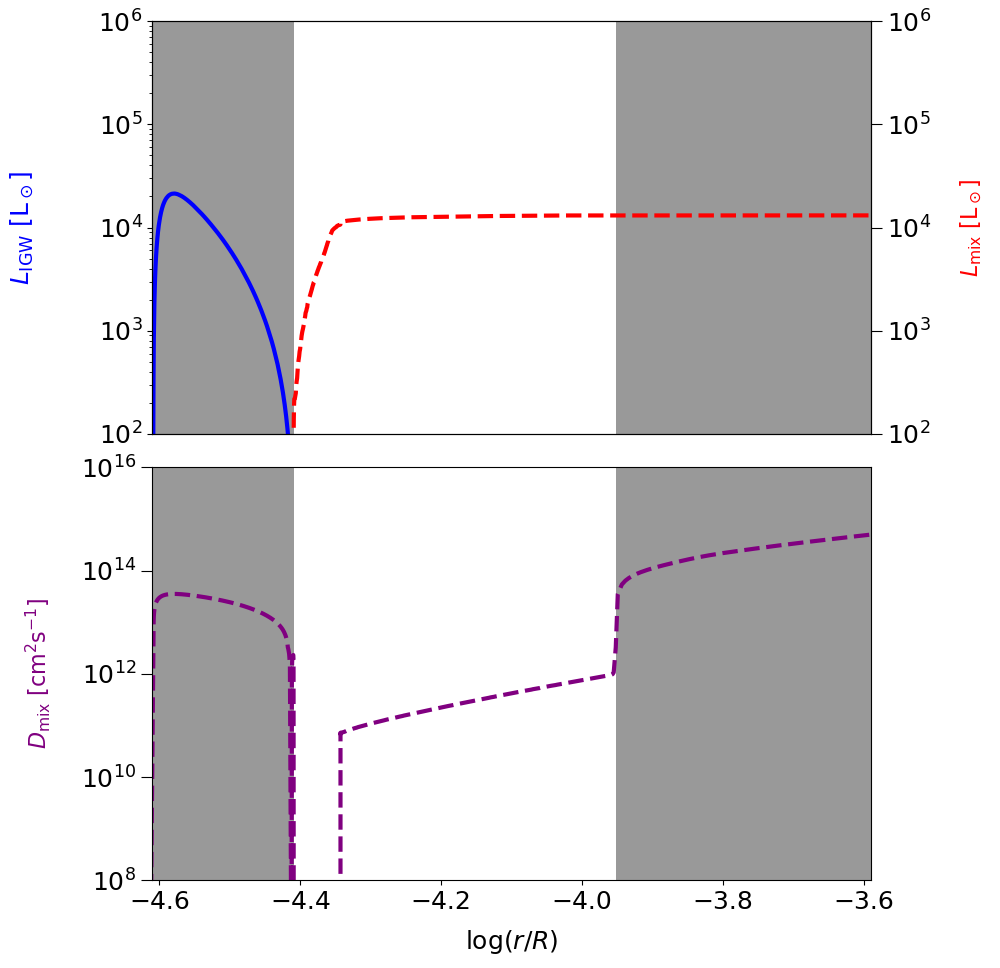}
\caption{Internal structure of a 4$\textit{M}_\odot$ star with $Z$=0.014 at the moment of He-shell ignition during TP-AGB stage. Here, $r$ is the radial distance from the core, and $R$ is the stellar radius. In upper panel, the values corresponding to the left vertical axis are represented by solid lines, while the dashed lines correspond to the values on the right vertical axis. The lower panel shows $D_{\text {mix }}$ in these zones. The gray shaded area represents the convective zone. $L_{\text {mix }}$ represents the power required to trigger mixing up to this zone. }
\label{fig:IGW}
\end{figure}

\begin{equation}
\int N^2 K \, dm,
\label{eq:1}
\end{equation}
Where $m$ is mass, $K$ is the thermal diffusion coefficient which given by
\begin{equation}
K = \frac{4acT^3}{3 \kappa \rho^2 C_P}=\frac{16\sigma T^3}{3 \kappa \rho^2 C_P},
\label{eq:2}
\end{equation}
where $\sigma$ = ac/4 is Stefan-Boltzmann constant, $T$ is tempature, $\kappa$ is opacity and $C_P$ is constant pressure specific heat capacity.

IGW is triggered by turbulent convection. Based on \citet{2013MNRAS.430.2363L} and \citet{2020ApJ...901L..18S} the wave luminosity of IGW in convective regions is given by
\begin{equation}
L_{\text{IGW}} = F_{\text{conv}} = M L_{\text{conv}}.
\label{eq:3}
\end{equation}
Here, $M$ is the convective Mach number. $L_{\text {conv }}$ means convective luminosity.
In order to sustain mixing within a region, the power of the IGW must exceed the threshold $L_{\text {mix }}$ required to maintain mixing in that region.

Fig.\ref{fig:IGW} depicts the internal structure of the star shortly after ignition in the He-shell. The shaded areas represent the convective regions. This includes the convective thermal pulse in the internal and the convective envelope in the external. A radiative zone exists between the convective thermal pulse and the convective envelope. Combine with Eq.(\ref{eq:1}), we take the top of the convective thermal pulse as the lower integration limit and perform a mass integral of $N^{2}K$. The position of the upper integration limit corresponds to the IGW power required to sustain mixing up to that region. When the upper limit is set at the bottom of the convective envelope, we derive the IGW power needed to maintain mixing across the radiative zone between the convective thermal pulse and the convective envelope. After normalizing the units of luminosity and power, this value is approximately $10^{4} \textit{L}_\odot$.

In Fig.\ref{fig:IGW}, it can be observed that during the TP-AGB stage of intermediate-mass stars, the $L_{\text {IGW }}$ in the He-shell region reaches $10^{4} \textit{L}_\odot$ to $10^{5} \textit{L}_\odot$ and exceeds the $L_{\text {mix }}$ value. This implies that IGW can induce mixing in the radiative zone between the convective thermal pulse and the convective envelope.
\subsubsection{Mixing coefficients setting}
The mixing induced by IGW requires specific mixing coefficients in each region. Schwab (\citeyear{2020ApJ...901L..18S}) applied fixed $D_{\text{mix}}$ = $10^{10} \, \text{cm}^2 \, \text{s}^{-1}$ and $D_{\text{mix}}$ = $10^{14} \, \text{cm}^2 \, \text{s}^{-1}$ in the context of mixing coefficients in low-mass stars. However, artificially setting the diffusion coefficient may not be accurate. In this work, we adopt the framework proposed by Herwig (\citeyear{2023MNRAS.525.1601H}). We use the mixing coefficient $D_{\text {mix }}$ generated by the mixing zone excited by IGW
\begin{equation}
\text{D}_{\text{IGW}} \approx \eta \cdot \frac{(\nabla \times \mathbf{u})^2 \cdot K}{N^2},
\label{eq:4}
\end{equation}
\begin{equation}
|\nabla \times \mathbf{u}| \propto L^{1/3}.
\label{eq:5}
\end{equation}
which u represents the convective velocity. The dimensionless coefficient $\eta$ is set to 0.1, with its value determined by 3D hydrodynamical simulations \citep{2016ApJ...821...49G}.

Similar to Schwab (\citeyear{2020ApJ...901L..18S})'s approach in modeling the first He-flash of low-mass stars, we set $L_{\mathrm{He}}$ = $10^{4} \textit{L}_\odot$ as the threshold for IGW mixing to confine its activation specifically during the He-shell ignition phase. This luminosity threshold is only exceeded transiently during the brief He-shell ignition event. We apply the IGW mixing in the regions where $L_{\mathrm{He}}$ \textgreater $10^{4} \textit{L}_\odot$.
\begin{figure*}[htbp]
    \centering
    \begin{minipage}[b]{0.45\textwidth}
        \centering
        \includegraphics[width=\textwidth]{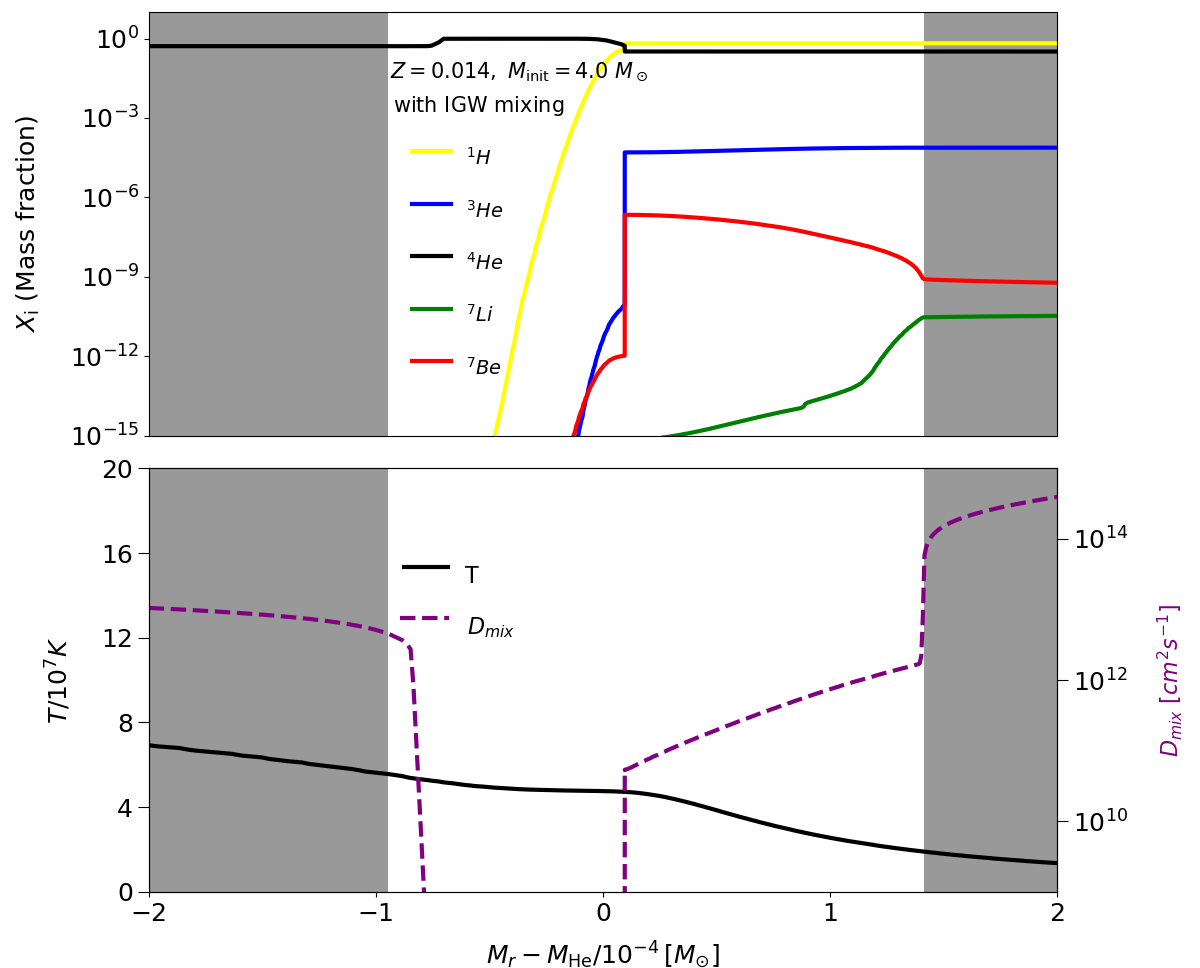}
    \end{minipage}
    \hfill
    \begin{minipage}[b]{0.45\textwidth}
        \centering
        \includegraphics[width=\textwidth]{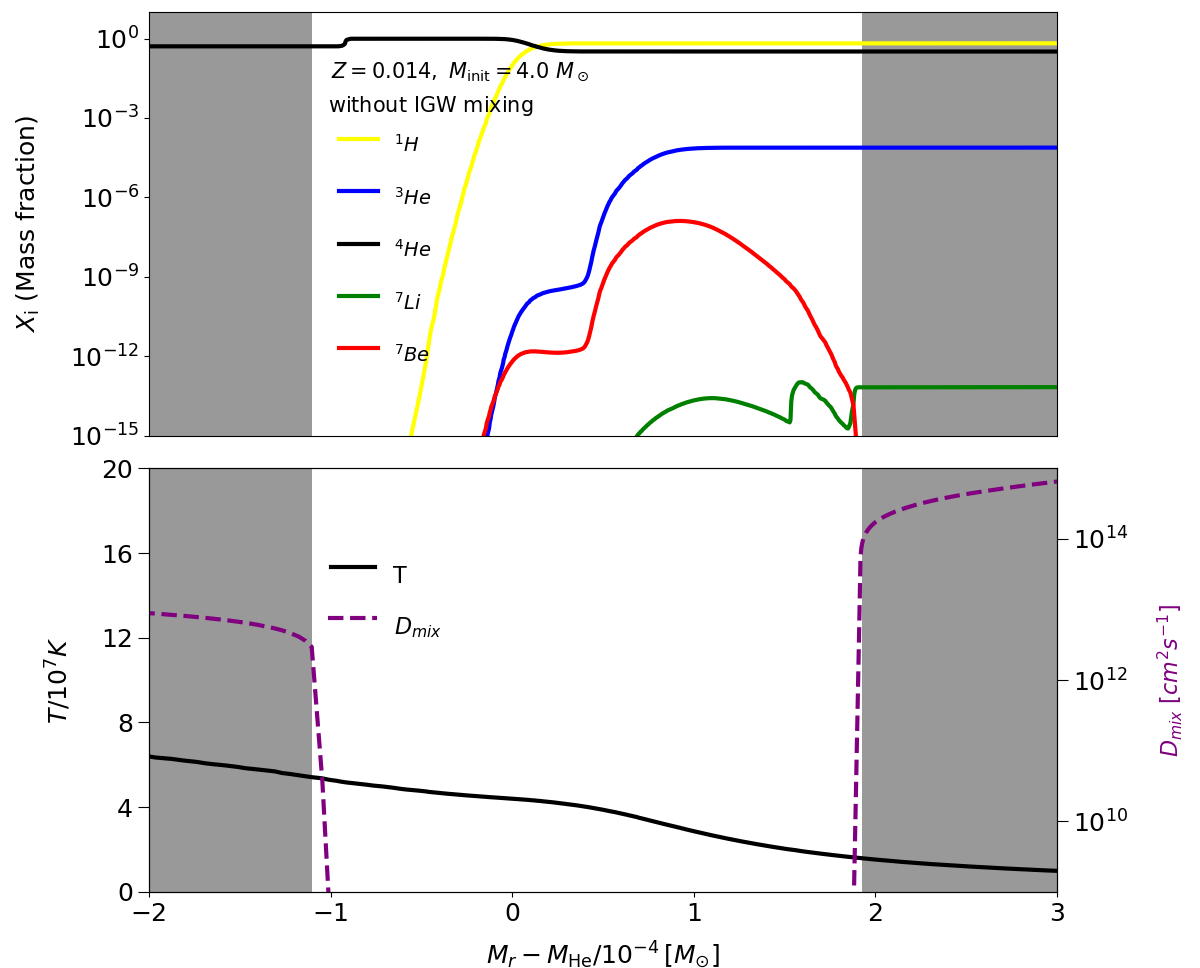}
    \end{minipage}
    \caption{Internal structure and elemental abundances from the convective thermal pulse to the convective envelope during AGB for 4${M}_\odot$ with $Z$=0.014. The gray shaded area represents the convective zone and the middle region is the convection interruption zone between the convective thermal pulse and the convective envelope. The left panel shows the case with IGW mixing considered but the right panel is the case without IGW mixing. For x-axis, $M_{\text{r}}$ represents the Lagrangian mass coordinate, $M_{\text{He}}$ means the Lagrangian mass coordinate of the outer boundary of the convective thermal pulse.}
    \label{fig:3}
\end{figure*}
\begin{figure*}[htbp]
    \centering
    \begin{minipage}[b]{0.45\textwidth}
        \centering
        \includegraphics[width=\textwidth]{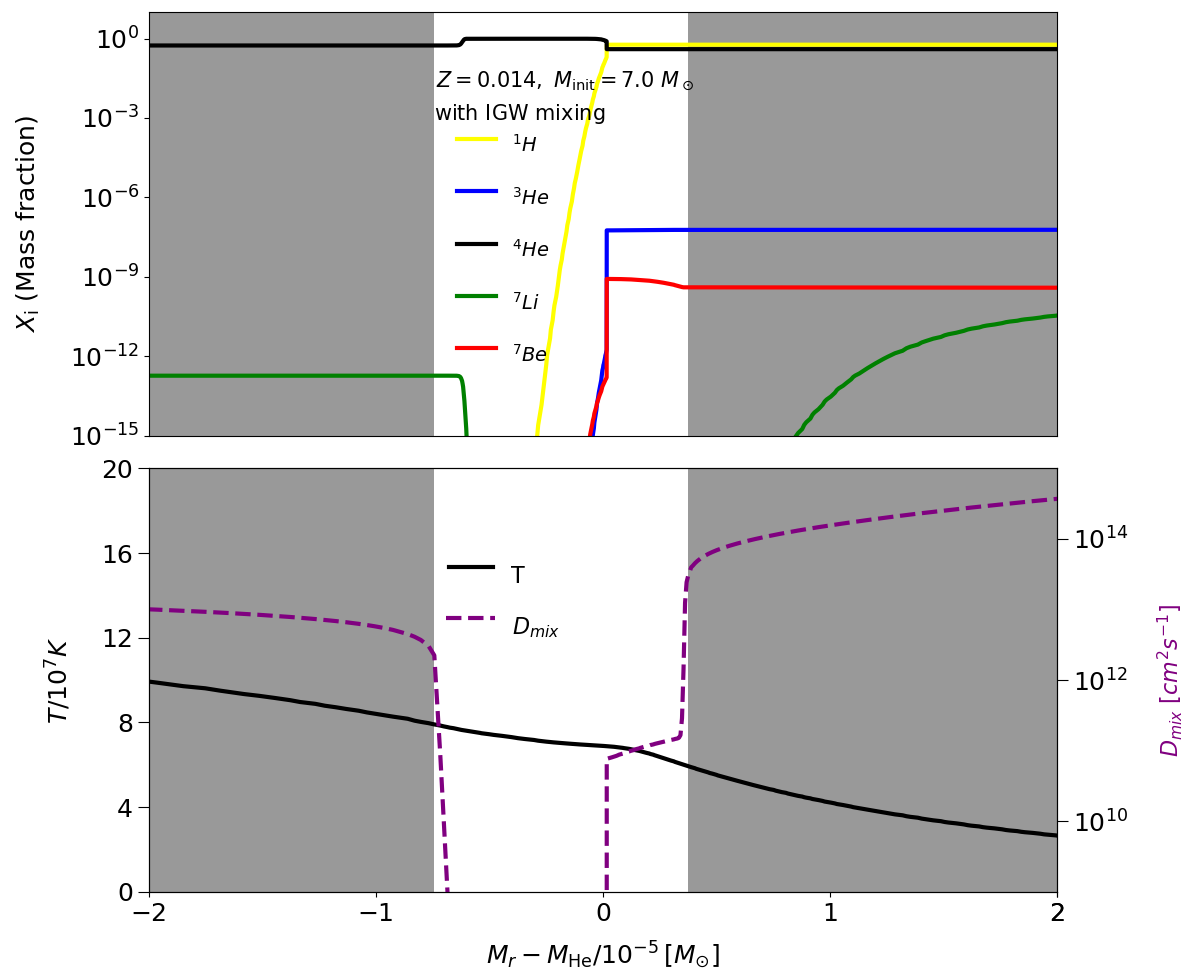}
    \end{minipage}
    \hfill
    \begin{minipage}[b]{0.45\textwidth}
        \centering
        \includegraphics[width=\textwidth]{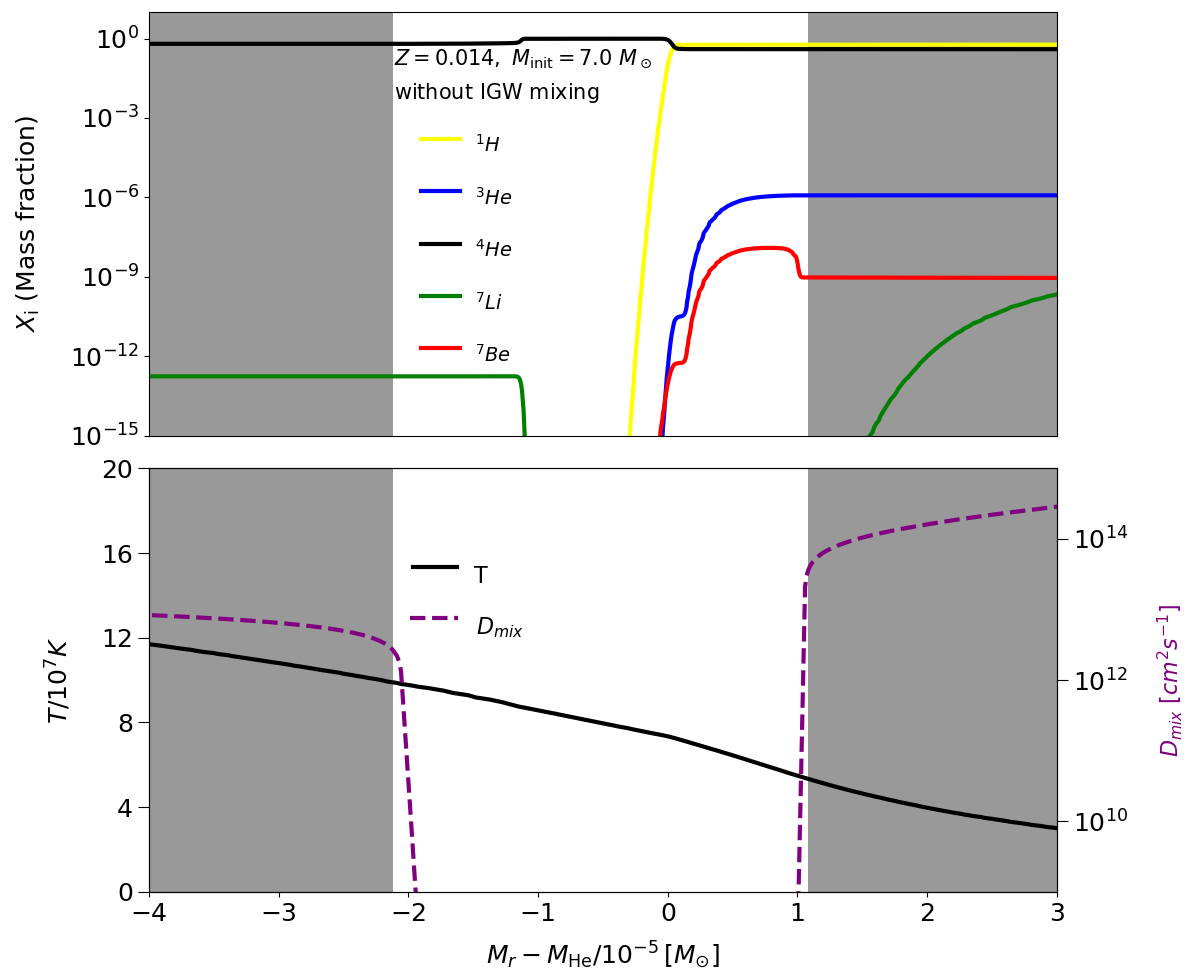}
    \end{minipage}
    \caption{Similar as Fig.3 but for 7$M_\odot$ models.}
    \label{fig:4}
\end{figure*}
Additionally, we consider element diffusion mechanism which is primarily driven by gravity sedimentation \citep{1994ApJ...421..828T,2018ApJS..234...34P}. In this paper, we focus on the diffusion of the 8 main elements during the AGB, including \ce{^1H}, \ce{^4He}, \ce{^7Be}, \ce{^7Li}, \ce{^{12}C}, \ce{^{13}C}, \ce{^{14}N}, and \ce{^{16}O}.
\section{Results}
We demonstrate the impact of IGW mixing mechanisms on Li production, present a Li yield grid, estimate the contribution rate of AGB stars to the Galactic Li, and compare with observational results.
\subsection{The impact of IGW mixing on Li abundance}
\begin{figure*}[htbp]
    \centering
    \begin{minipage}[b]{0.32\textwidth}
        \centering
        \includegraphics[width=\textwidth]{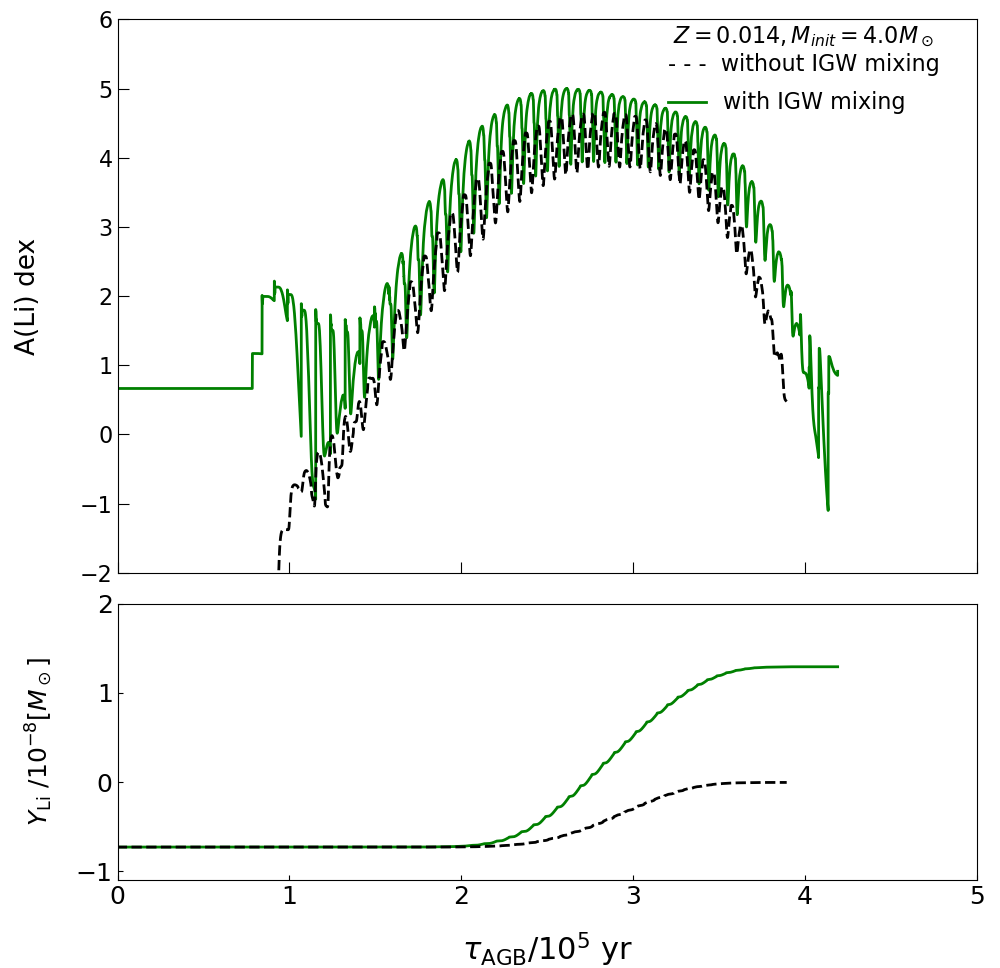}
    \end{minipage}
    \hfill
    \begin{minipage}[b]{0.32\textwidth}
        \centering
        \includegraphics[width=\textwidth]{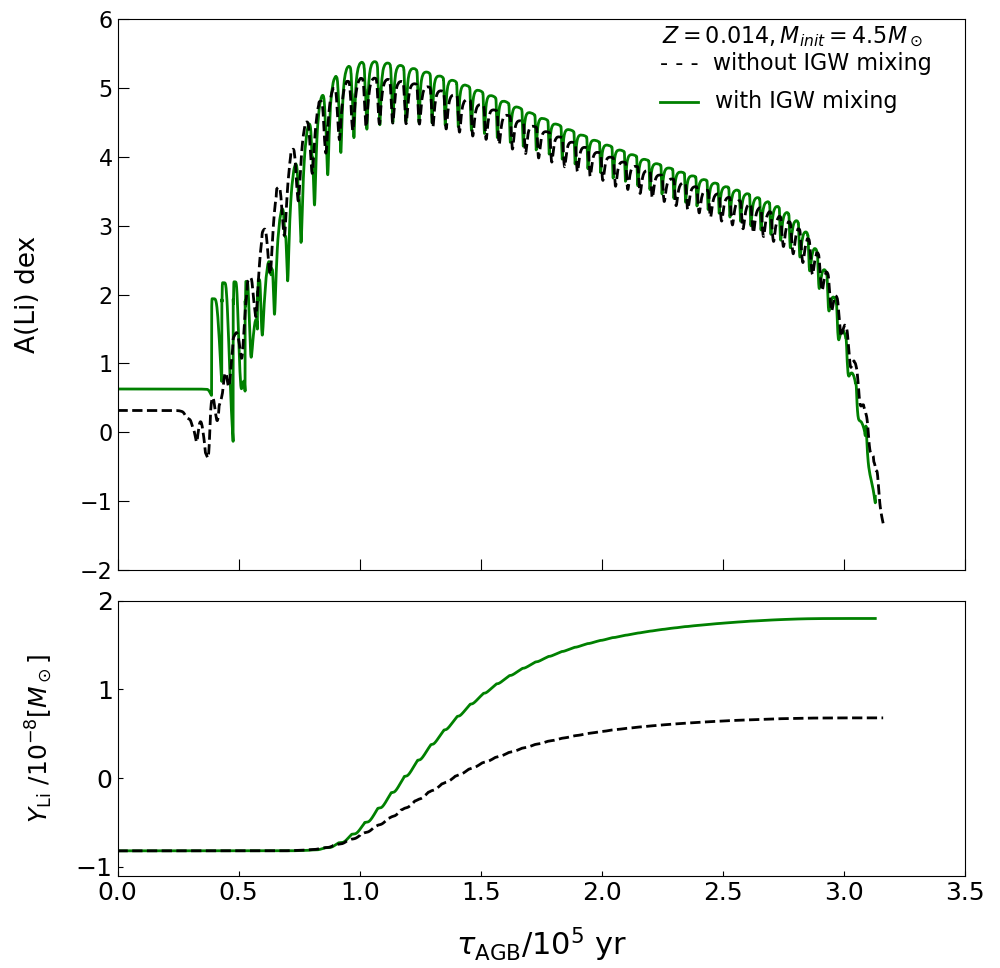}
    \end{minipage}
    \hfill
    \begin{minipage}[b]{0.32\textwidth}
        \centering
        \includegraphics[width=\textwidth]{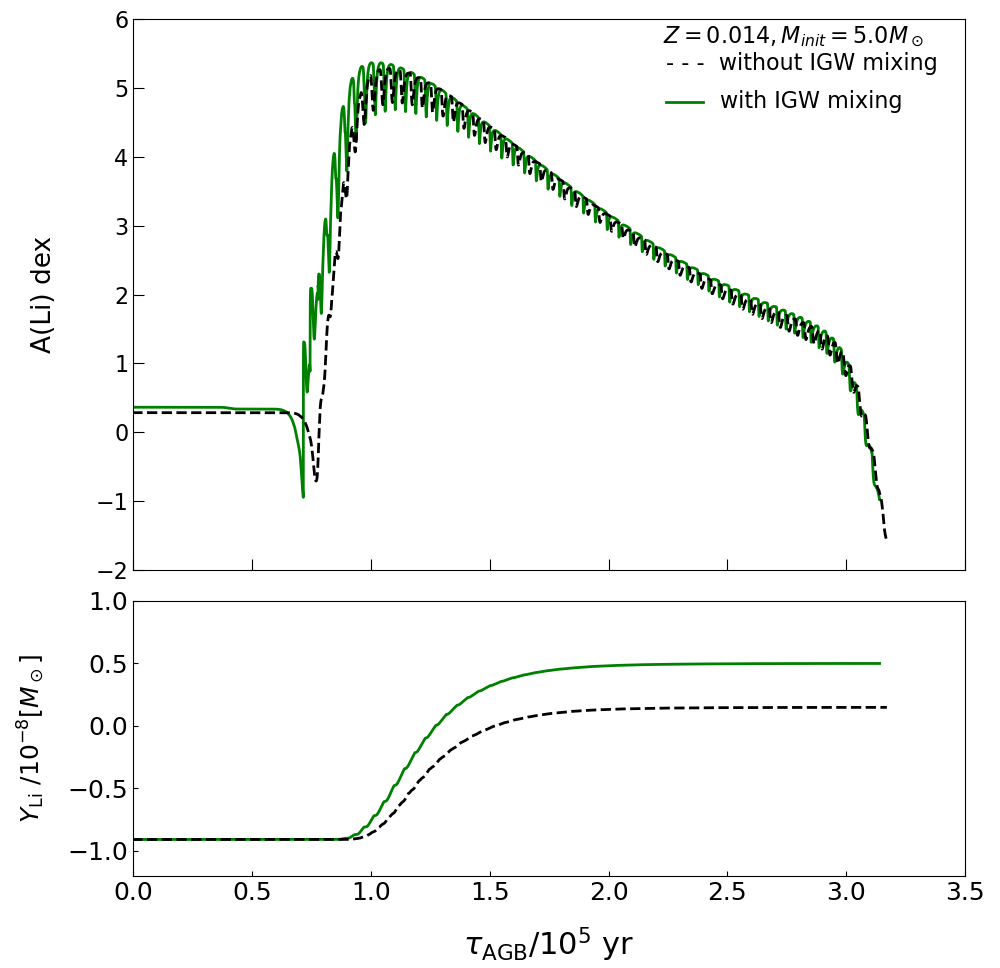}
    \end{minipage}
    \vspace{0.5em}%
    \begin{minipage}[b]{0.32\textwidth}
        \centering
        \includegraphics[width=\textwidth]{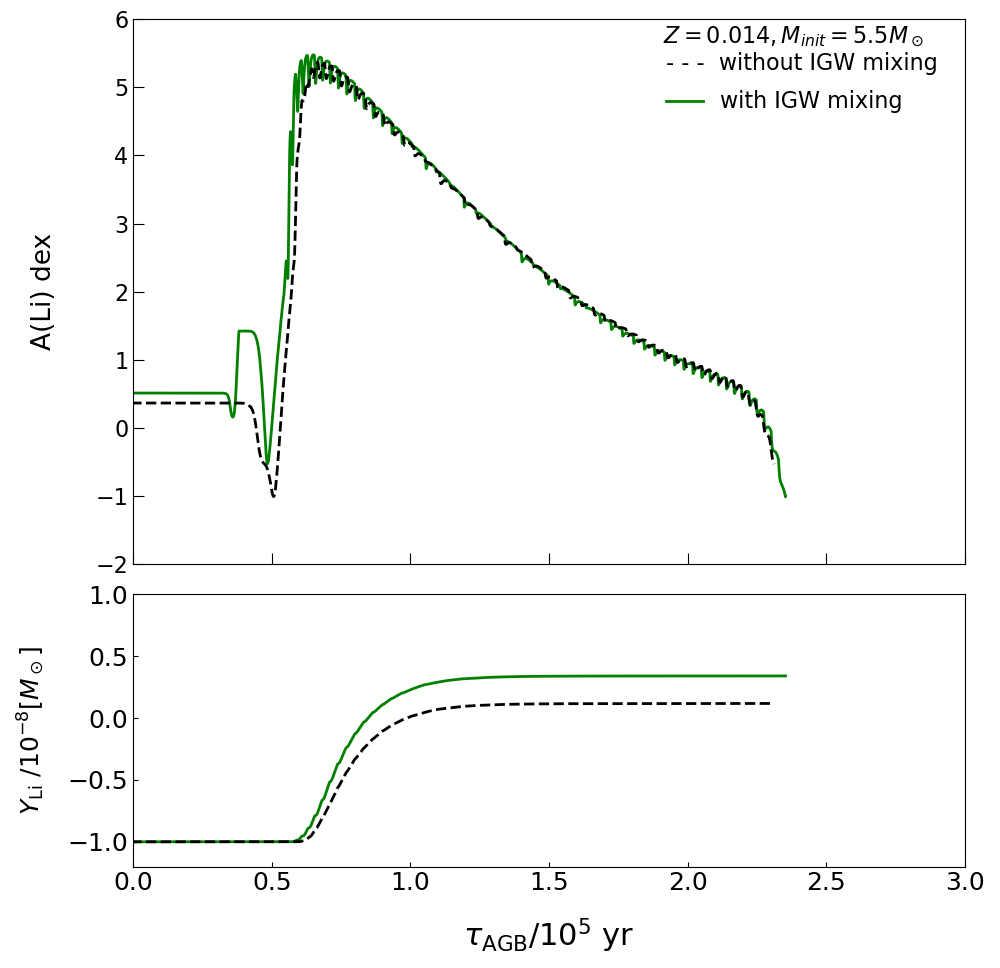}
    \end{minipage}
    \hfill
    \begin{minipage}[b]{0.32\textwidth}
        \centering
        \includegraphics[width=\textwidth]{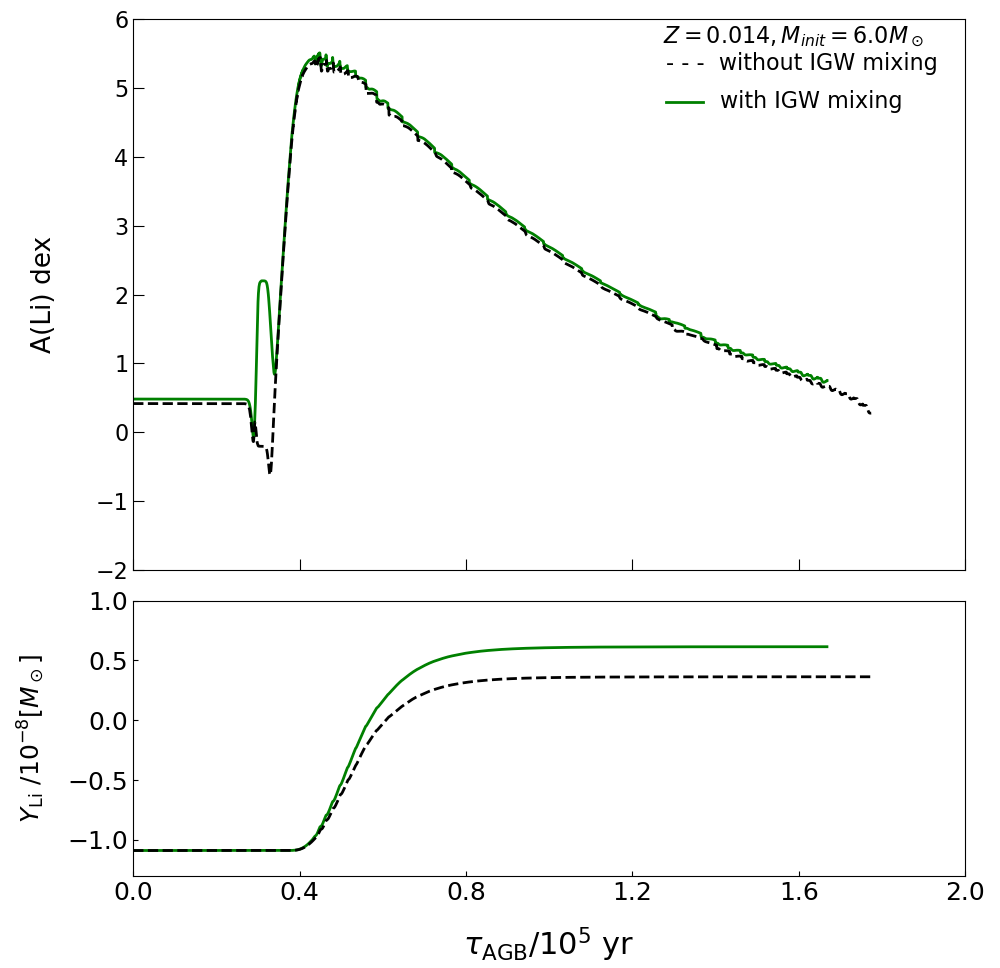}
    \end{minipage}
    \hfill
    \begin{minipage}[b]{0.32\textwidth}
        \centering
        \includegraphics[width=\textwidth]{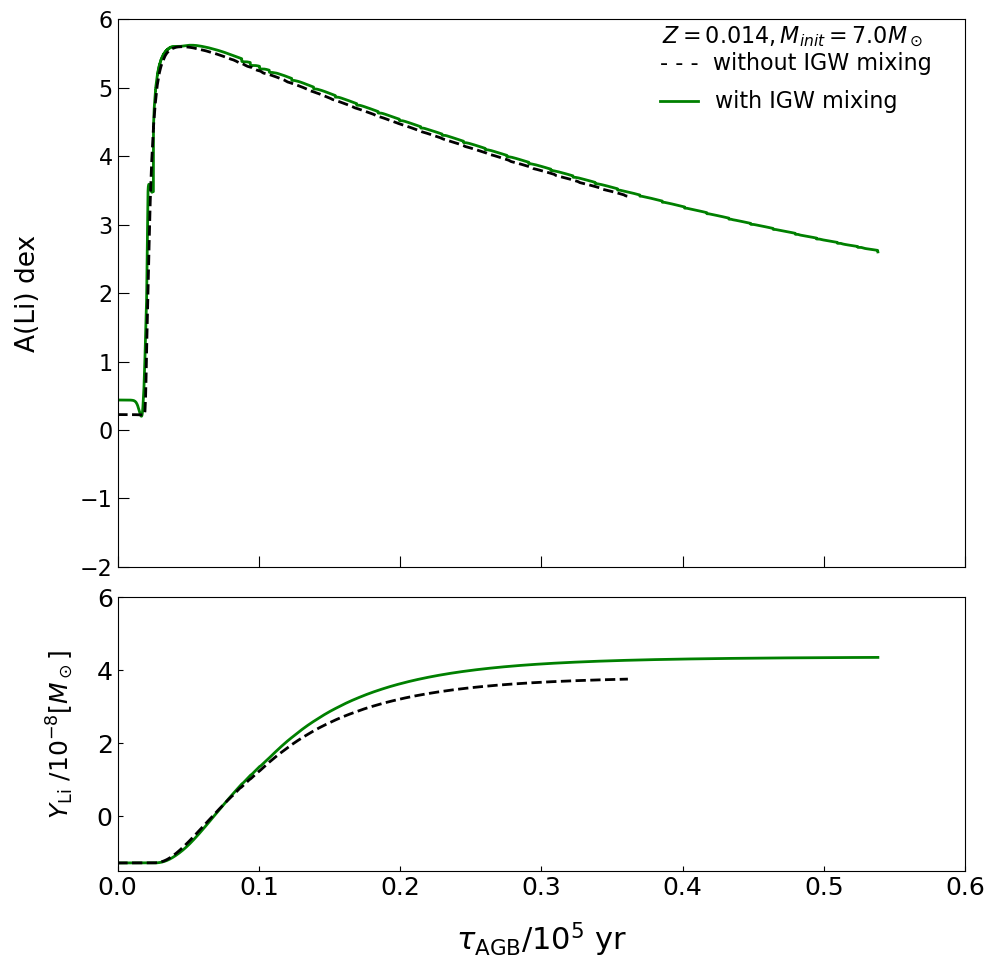}
    \end{minipage}
    \caption{Variation of \ce{^7Li} abundance and total Li yields $Y_{\rm{Li}}$ (subtracting the initial Li mass) over time during AGB stage for stars of different masses under solar metallicity. We use dashed lines to represent the case without IGW mixing, and the meaning of the x-axis is the same as in Fig.1}
    \label{fig:5}
\end{figure*}
Fig.\ref{fig:3} shows a comparison of models with and without considering IGW mixing for a star with initial mass of 4$M_\odot$ and $Z$=0.014 by the element abundances, mixing coefficients, temperatures, and convective zones near the He-shell during the TP-AGB phase. Without IGW mixing, the mixing in the non-convective regions hardly occurs because of too low $D_{\text {mix }}$. Even when considering convective overshooting effects, the mixing region can only expand slightly. As shown by red line in Fig.\ref{fig:3}, a large amount of \ce{^7Be} produced in the radiative zone between the convective thermal pulse and the convective envelope. Based on the CF mechanism \citep{1955ApJ...122..271F,1971ApJ...164..111C}, the region around $4 \sim 8 \times10^{7}\mathrm{K}$ is the primary production zone for \ce{^7Be}. Without IGW mixing, a significant portion of this region located below the convective envelope, thus a substantial amount of internal \ce{^7Be} from this region can not be transported into the convective envelope. This is particularly evident in stars which initial mass just satisfies the requirements for the HBB mechanism. Below the convective envelope, there still exists a region where the temperature is suitable for the $^{3}\text{He}(\alpha{},{} \gamma)^{7}\text{Be}(e, \nu)^{7}\text{Li}$ reaction and where the abundance of \ce{^3He} is relatively high. This is precisely why this region is enriched with \ce{^7Be}. In the models with IGW mixing, the mixing in the non-convective regions can also efficiently work because $D_{\text {mix }}$ is higher than $10^{10} \, \text{cm}^2 \, \text{s}^{-1}$ in these regions.

Similarly, Fig.\ref{fig:4} shows the internal structure for 7$M_\odot$ star. Although the mixing induced by IGW in 7$M_\odot$ model is similar to that in 4$M_\odot$ model, the Li production in the 7$M_\odot$ model with IGW mixing remain comparable to those without IGW mixing. This occurs because the bottom temperature of the convective envelope in the 7$M_\odot$ model exceeds $4 \times 10^{7}$K, satisfying the conditions for \ce{^7Be} production and enabling efficient Li synthesis even without IGW mixing. For a star with 4$M_\odot$, the mass of IGW mixing region below the convective envelope is approximately $2 \times 10^{-4}$$M_\odot$, the mass of \ce{^7Be} is about $2 \times 10^{-11}$$M_\odot$. In the case of 7$M_\odot$, the mass of \ce{^7Be} below the convective zone is only about $10^{-13}$$M_\odot$. Furthermore, the effectiveness of the IGW mixing decreases continuously with an increasing number of pulses. Considering that larger mass stars require more pulses for the He-shell to achieve higher Helium-luminosity, internal \ce{^3He} may be depleted before IGW mixing effectively operates. On the other hand, larger initial mass stars have higher temperatures at the bottom of their convective envelope, making the HBB mechanism more effective. Therefore, IGW mixing does not significantly enhance the production of Li in higher initial mass stars.
\begin{figure*}[htbp]
    \centering
    \begin{minipage}[b]{0.96\textwidth}
        \centering
        \includegraphics[width=\textwidth]{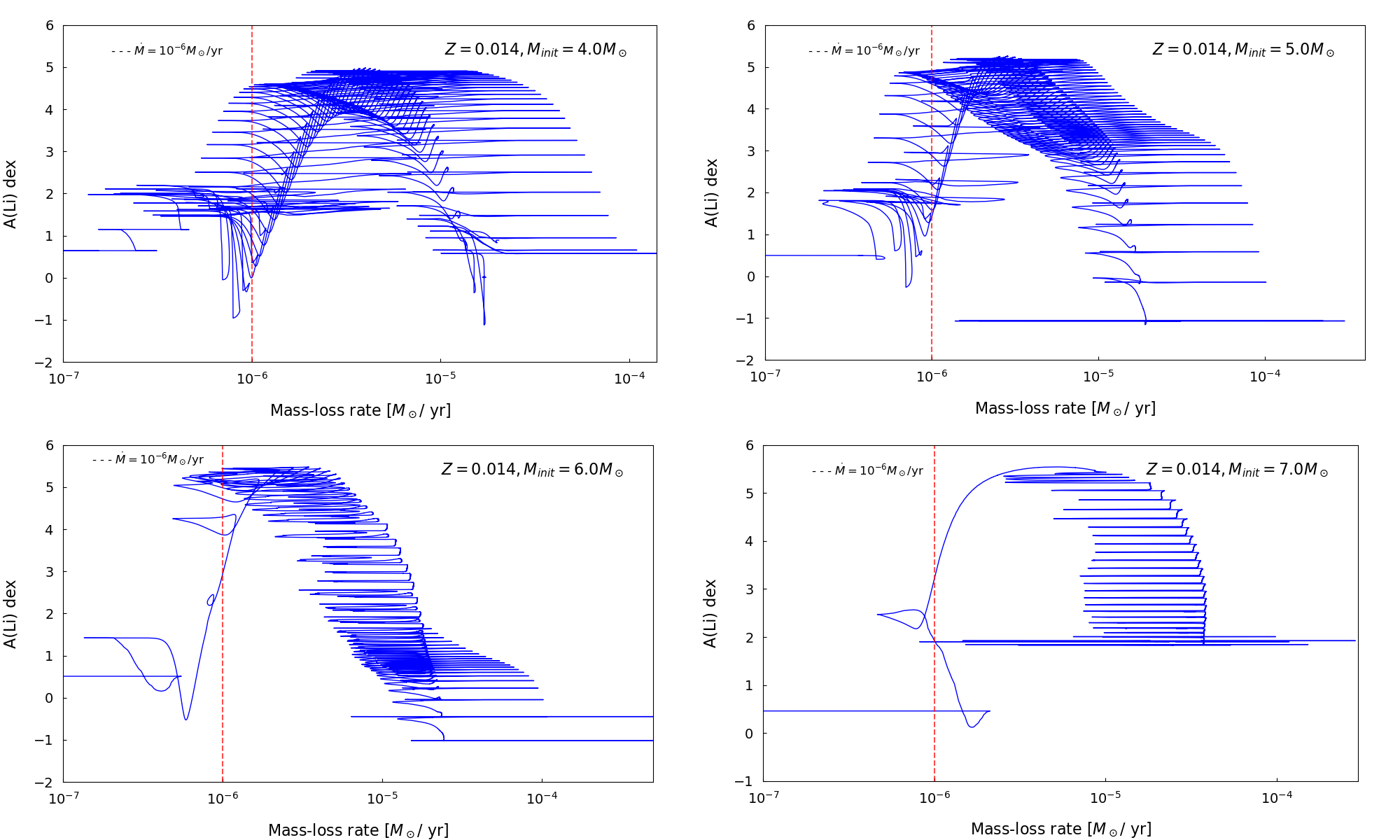}
    \end{minipage}
    \caption{Variation of surface \ce{^7Li} abundance (y-axis) and mass-loss rates (x-axis) during TP-AGB phase for solar metallicity stars with different initial mass in our model. We have indicated the position with a mass-loss rate of $10^{-6}~M_{\odot}$yr$^{-1}$ using a red dashed line.}
    \label{fig:6}
\end{figure*}

Fig.\ref{fig:5} compares the temporal evolution of surface \ce{^7Li} abundances and cumulative yields (calculated via Eq.(\ref{eq:9}) and Eq.(\ref{eq:10})) for $4\sim7$$M_\odot$ stars with and without IGW mixing throughout the entire TP-AGB phase. In our model, the peak value of A(Li) during the AGB phase exceeds 5. This result is consistent with \citet{2016ApJ...825...26K}, which prescribed a fixed mixed mass below the convective envelope in their AGB models. Based on our simulations, IGW mixing is likely this extra-mixing assumed by \citet{2016ApJ...825...26K}. Since the effect of IGW mixing is more efficient in lower-mass stars, the Li enhancement effect is most significant in stars whose initial mass just meets the critical value for the HBB mechanism. We also can find that regardless of whether the IGW mixing mechanism is considered or not, the peak in surface \ce{^7Li} abundance occurs during the early pulses. While for cases involving IGW-mixing, the Li abundance during the same evolutionary phase is higher than that in the models without IGW. Although the difference remains below 1 dex, the logarithmic nature of the A(Li) scale implies that even a modest difference in abundance corresponds to a severalfold enhancement in the mass fraction of Li at the stellar surface and in stellar winds. Apart from the 7$M_\odot$ model, with IGW mixing, the Li production increases by several times to an order of magnitude compared to scenarios without IGW mixing. Since stellar mass ejection primarily occurs during AGB, achieving a positive Li yield requires that the A(Li) during AGB exceeds the initial value by at least a certain amount and persists for a specific duration. Taking the models with $Z$=0.014 as an example, combined with the initial A(Li) value of 3.26, the total Li mass ejected by the star over its entire life cycle must exceed $7.27 \times 10^{-9}$$M_\odot$. Otherwise, the Li production would be insufficient to compensate for the Li consumed from the ISM during the star formation.
\subsection{Comparison with observational samples}
\begin{figure*}[htbp]
    \centering
    \begin{minipage}[t]{0.48\textwidth} 
        \centering
        \includegraphics[width=\linewidth]{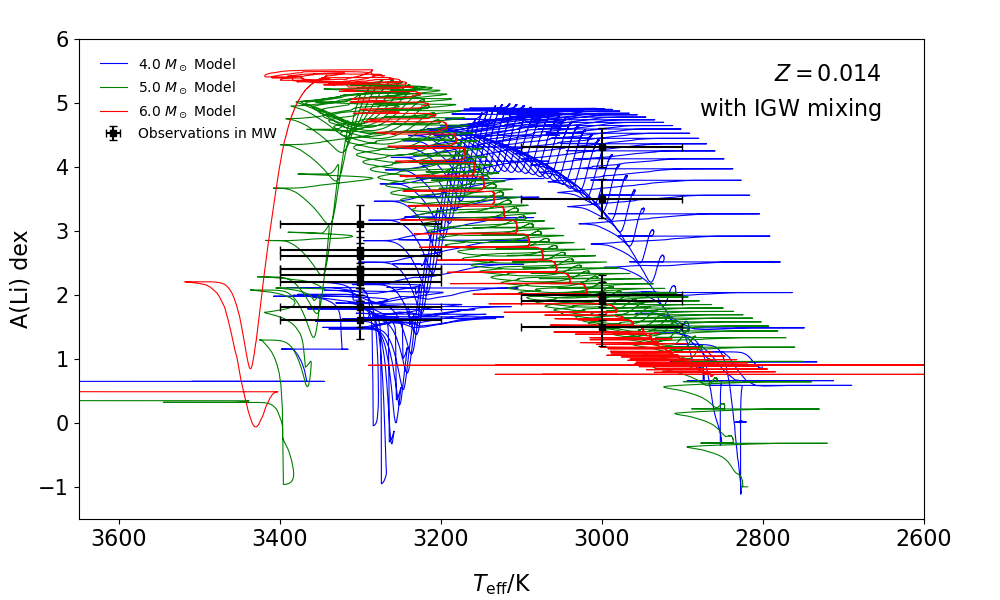} 
        \vspace{1ex} 
    \end{minipage}
    \hfill
    \begin{minipage}[t]{0.465\textwidth}
        \centering
        \includegraphics[width=\linewidth]{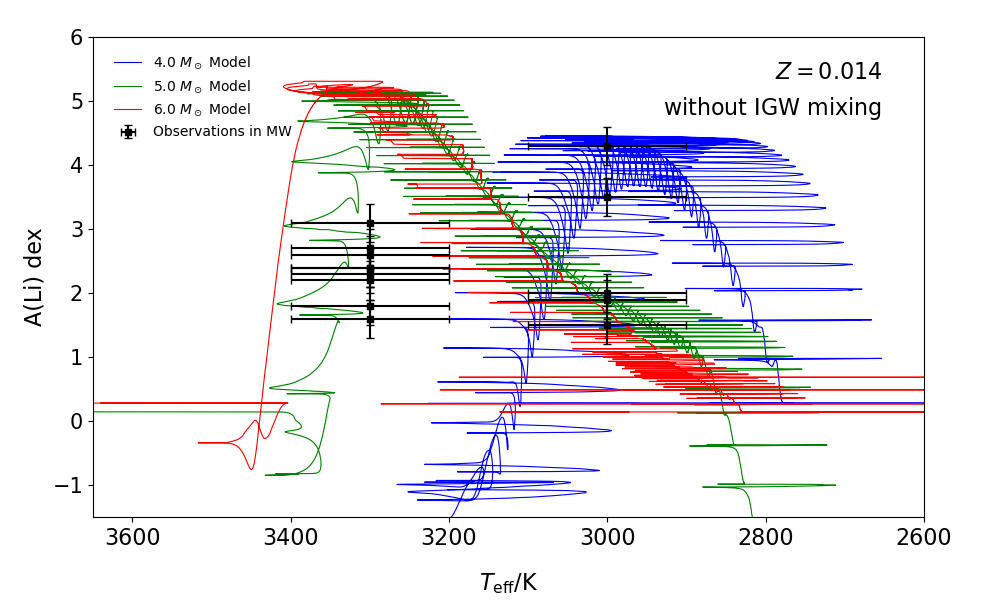}
        \vspace{1ex}
    \end{minipage}
    \vspace{1em}
    \begin{minipage}[t]{0.48\textwidth}
        \centering
        \includegraphics[width=\linewidth]{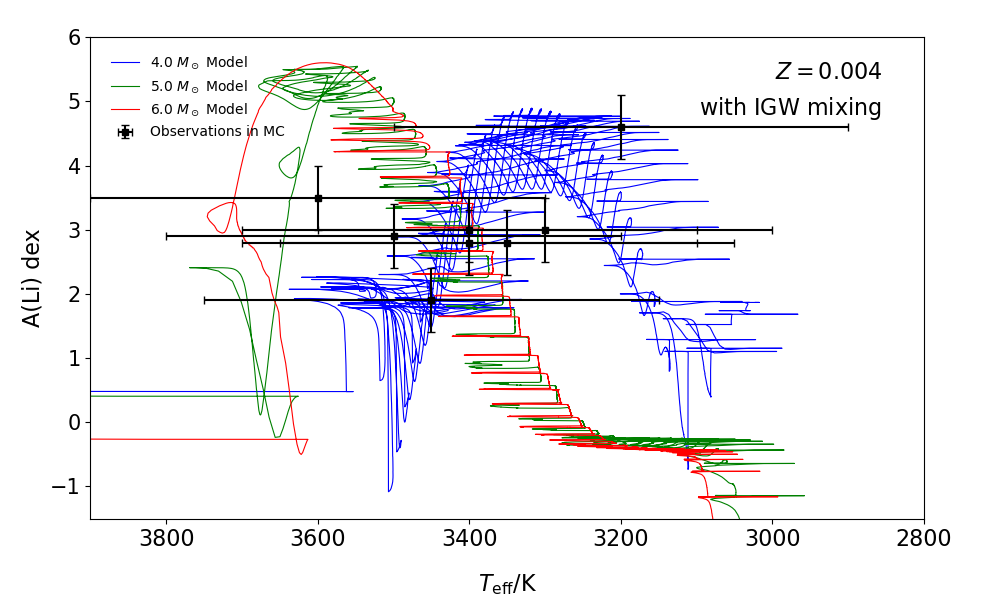}
        \vspace{1ex}
    \end{minipage}
    \hfill
    \begin{minipage}[t]{0.47\textwidth}
        \centering
        \includegraphics[width=\linewidth]{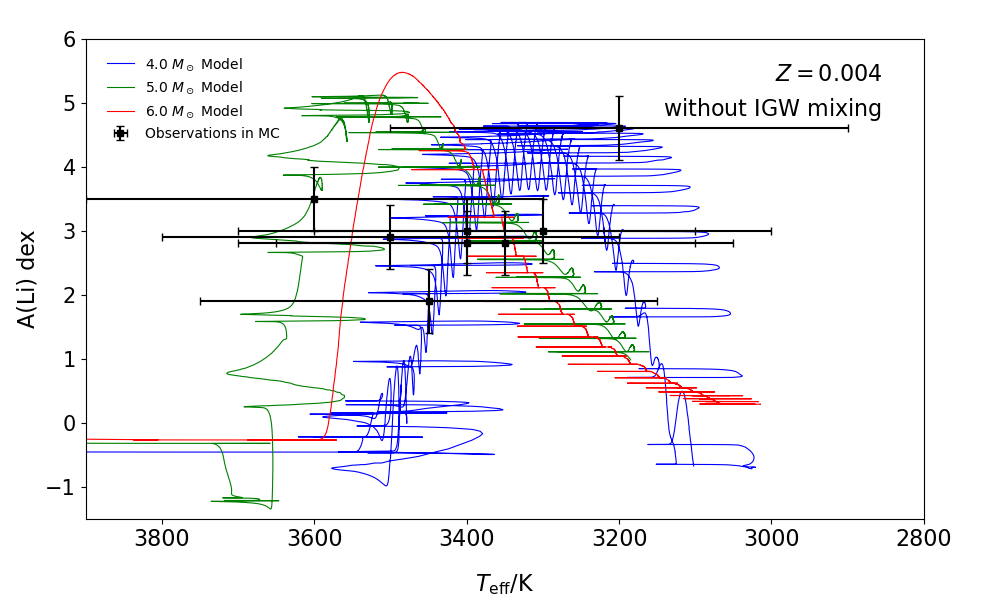}
        \vspace{1ex}
    \end{minipage}
    \caption{Comparison of observed and modeled Li-rich AGB samples in the Milky Way and the Magellanic Clouds. The upper panel shows the Milky Way sample with effective temperature versus \ce{^7Li} abundance, with $T_{\text{eff}}$ errors of $\pm 100 \text{K}$, \ce{^7Li} abundance errors of $\pm 0.3 \text{dex}$. The lower panel shows the Magellanic Clouds sample with $T_{\text{eff}}$ errors of $\pm 300 \text{K}$, \ce{^7Li} abundance errors of $\pm 0.5 \text{dex}$. The models with IGW mixing are all displayed on the left, while the models without IGW mixing are displayed on the right.} 
    \label{fig:7}
\end{figure*}

Based on current observations, there are about 22 Li-rich AGB stars, with 14 located in the Milky Way \citep{2013A&A...555L...3G,2019A&A...623A.151P} and 8 in the Magellanic Cloud \citep{1983ApJ...272...99W,1989ApJ...345L..75S,1990ApJ...361L..69S,1993ApJ...418..812P,2011A&A...531A..88U}. Their Li abundances, A(Li) are all above 1.5 dex, with R Nor \citep{2011A&A...531A..88U} and R Cen \citep{2019A&A...623A.151P} reaching 4.6 dex and 4.3 dex, respectively. All these stars are O-rich and exhibit s-process element enrichment, suggesting the occurrence of HBB. Based on the observations of the Rb~\text{I} 7800~\AA\ absorption line, the mass-loss rates of the samples in the Milky Way are all less than $10^{-6}$$M_{\odot}$yr$^{-1}$. When the mass-loss rates exceeds $10^{-6}$$M_\odot$yr$^{-1}$, the Rb~\text{I} absorption lines in the Milky Way observed samples become excessively strong \citep{2014A&A...564L...4Z,2017A&A...606A..20P}. Fig.\ref{fig:6} shows the evolution of A(Li) on the stellar surface with mass-loss rates. During the AGB phase, the star's mass-loss rates increases with each pulse oscillation, accompanied by an increase in Li abundance due to the HBB mechanism and a decrease in Li abundance after the depletion of \ce{^3He}. In our model, after entering the AGB phase, there is a distinct stage where the mass-loss rates remains below $10^{-6}$$M_\odot$yr$^{-1}$. Moreover, under reasonable mass-loss rates, there can exist a super Li-rich phase with A(Li) > 3.

Fig.\ref{fig:7} shows the comparison of our simulation results with observational data for both the Milky Way and the Magellanic Clouds at different metallicities. During the TP-AGB phase, as the number of pulses and time increase, the star's effective temperature oscillates and gradually decreases. Therefore, the effective temperature can effectively reflect the evolutionary progress of the star during the TP-AGB phase. Our results with or without IGW mixing can
cover observational samples well.

\subsection{Li yields for AGB stars}
We use the following calculation method to calculate Li yields
\begin{equation}
M_{\text{Li}} = M_{\text{Li}}^{\text{end}} - M_{\text{Li}}^{\text{ini}},
\label{eq:9}
\end{equation}
and
\begin{equation}
M_{\text{Li}}^{\text{end}} = \int_0^{\tau} X_{\text{Li}}(t) \cdot \dot{M} \, dt.
\label{eq:10}
\end{equation}
Here, \( M_{\text{Li}}^{\text{ini}} \) represents the mass of Li in the star at the initial moment, \( X_{\text{Li}}(t) \) is the surface Li abundance at time $t$. $\dot{M}$ is mass loss rate and $\tau$ is the moment marking the end of the AGB phase of the star.
Tab.\ref{tab:1} gives the results of the net production of Li after the AGB phase for stars with metallicities $Z$=0.014, $Z$=0.004, $Z$=0.0014, and $Z$=0.00014, initial masses ranging from 3.5$M_\odot$ to 7.5$M_\odot$.
Based on a standard stellar evolution model, majority of Li in the whole star is consumed when the star undergoes the first dredge-up \citep{2019MNRAS.484.2000D,2022A&A...668A.126G}. Therefore, stars are considered as a negative contribution to Li in ISM. However, based on our results shown in Tab.\ref{tab:1}, the positive contribution to Li in ISM may appear when the stars have enough massive to undergo the HBB during TP-AGB stage. We find that stars with an initial mass just meeting the HBB condition have a relatively high Li production. Li production initially decreases with increasing stellar mass but rises again for stars above 6$M_{\odot}$, this trend being consistent with previous findings \citep{2010MNRAS.402L..72V}. For stars with lower initial masses, IGW mixing is more effective. This not only the higher temperature at the bottom of the convective envelope overwhelms the IGW mixing effect in higher-mass stars, but also because lower-mass stars attain higher Helium-luminosities during the He-shell ignition phase of the TP-AGB. When a star has an initial mass higher than 6$M_\odot$, the temperature at the bottom of the convective envelope is high enough so that \ce{^7Be} can directly be produced. Therefore, the Li yields rise again. Stars with initial masses in the $3.5\sim6$$M_{\odot}$ range exhibit a higher birth rate compared to stars with initial masses above 6$M_{\odot}$, and their Li production contributes significantly to the ISM. Overall, the positive impact of IGW mixing generally decreases with increasing stellar mass and might be more pronounced in low-mass stars. Although the IGW mixing mechanism is more pronounced during the RGB and RC phases of low-mass stars, the limited duration of mixing relative to the total evolutionary phases and the low mass-loss rates during these stages may result in insufficient yields to significantly impact the composition of the ISM \citep{2022A&A...668A.126G}.

In the last column of Tab.\ref{tab:1}, we show the influence of IGW on the Li yields. Since some models exhibit negative yields, we display difference instead of ratio. We find the positive impact of IGW mixing decreases with increasing initial mass, particularly in terms of the enhacement magnitude for models without IGW mixing. We also note that in the case of low-metallicity models, the yields of some non-IGW models exceed those of the IGW models. This could be attributed to the higher temperature at the bottom of the convective envelope in metal-poor stars, which even surpasses $8\times 10^{7}$K. Under such conditions, the mixing applied in the non-convective zone between the convective thermal pulse and the convective envelope actually transports \ce{^7Be} from the convective envelope into hotter interior regions where \ce{^7Be} gets depleted.
\begin{figure*}[htbp]
\begin{minipage}[t]{0.48\textwidth} 
        \centering
        \includegraphics[width=\linewidth]{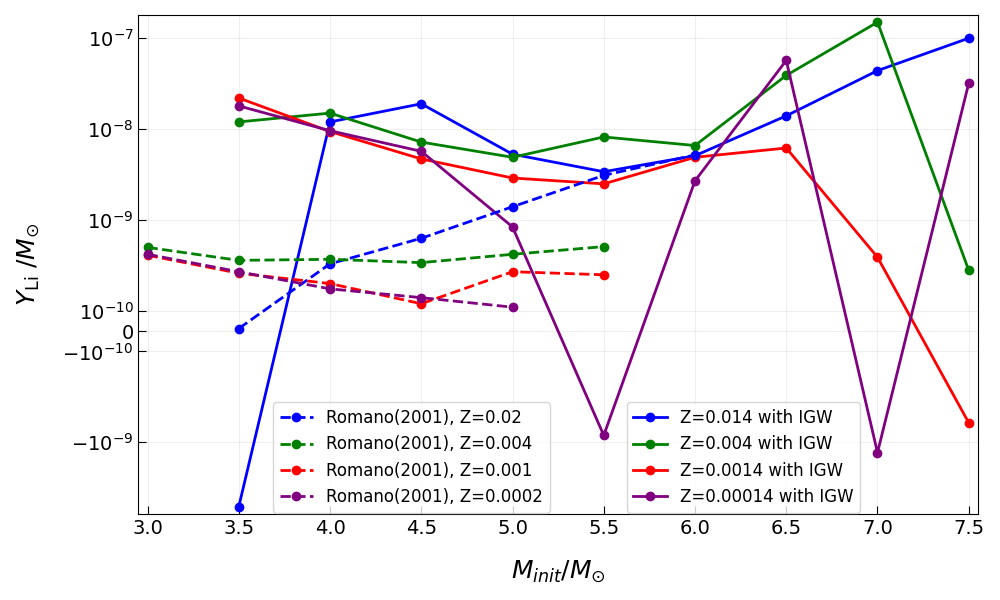} 
        \vspace{1ex} 
    \end{minipage}
    \hfill
    \begin{minipage}[t]{0.48\textwidth}
        \centering
        \includegraphics[width=\linewidth]{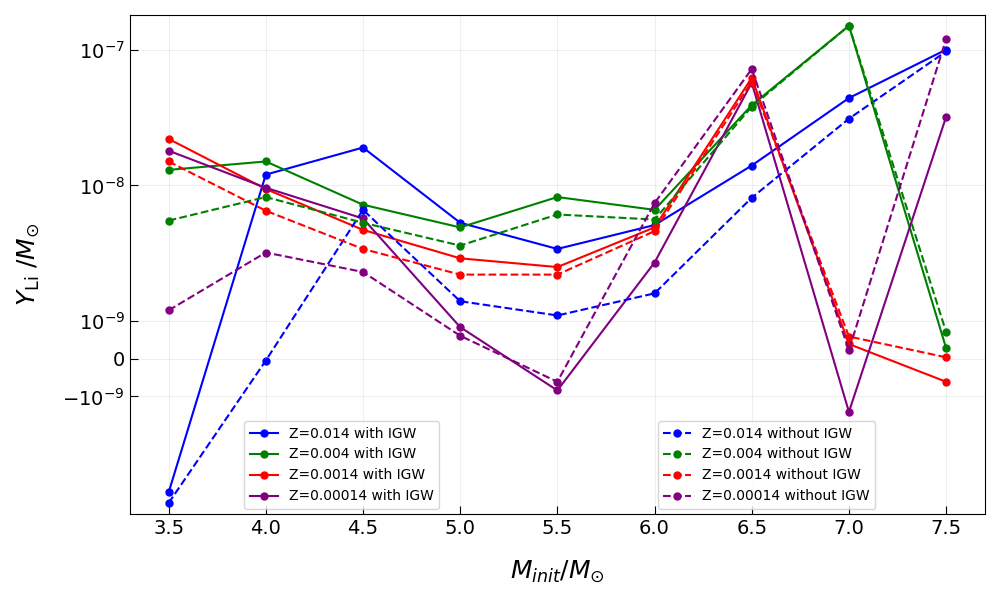}
        \vspace{1ex}
    \end{minipage}
    \caption{Li yields $Y_{\rm{Li}}$ for AGB stars with different initial masses and metallicities. In the left panel, the results with IGW mixing are given by colorful solid lines, while those without IGW mixing are showed by colorful dashed lines. In the right panel, the results in this work are given by colorful solid lines, while those in others are showed by colorful dashed lines.}
    \label{fig:8}
\end{figure*}
\begin{figure}[htbp]
\centering
\includegraphics[width=0.5\textwidth]{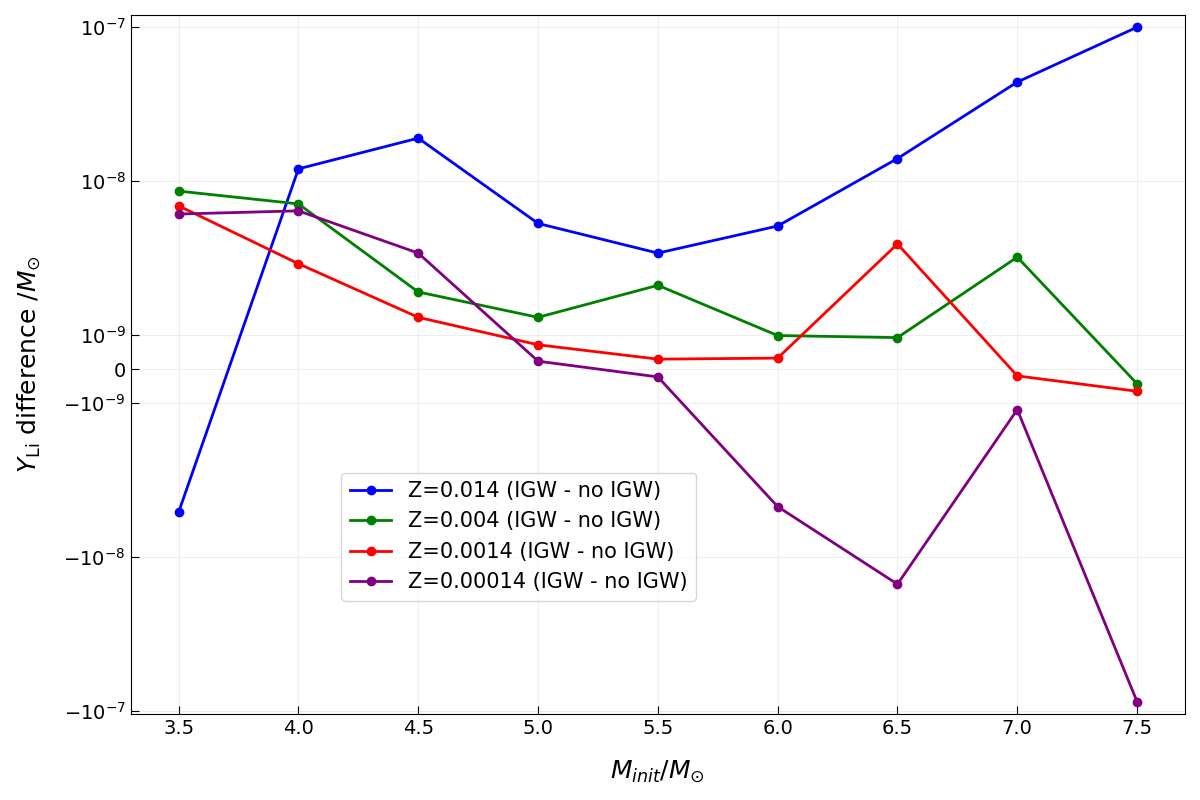}
\caption{Li yield $Y_{\rm{Li}}$ differences (y-axis) with and without IGW mixing under different metallicities and initial mass (x-axis) conditions }
\label{fig:9}
\end{figure}
In the previous investigations, Li yields produced by AGB stars have been calculated, such as Venture (\citeyear{2000A&A...363..605V}), Romano (\citeyear{2001A&A...374..646R}).
Fig.\ref{fig:8} gives the comparison between our work and theirs, as well as between our IGW model and the model without IGW mixing. Fig.\ref{fig:9} displays the Li yield gap with and without IGW mixing across different metallicities and initial masses in our models, using data from the last column of Tab.\ref{tab:1}. Obviously, comparative analysis between models with and without IGW mixing reveals that the influence of IGW mixing becomes more pronounced at lower initial masses. For systems with reduced metallicity, the overall trend exhibits a slight shift towards even lower mass regimes. In our model, with similar initial mass and metallicity, Li yields in our models are an order of magnitude higher than those in Venture (\citeyear{2000A&A...363..605V}) and Romano (\citeyear{2001A&A...374..646R}). Especially, in the models with a initial mass of 3.5-5$M_\odot$, the former is one order of magnitude higher than the later. The extra-mixing triggered by IGW play a non-negligible role, which is illuminated by Figs.3-5. Certainly, the yields from some of our model without IGW mixing also exceed those of previous results. This discrepancy could be attributed to revisions in the $^{7}\text{Be}(e, \nu)^{7}\text{Li}$ nuclear reaction rates used by this work, where the Simonucci (\citeyear{2013ApJ...764..118S}) reaction rates were adopted. In previous models of Venture (\citeyear{2000A&A...363..605V}) and Romano (\citeyear{2001A&A...374..646R}) , the nuclear reaction rates significantly underestimated the electron capture rates at temperatures below $10^7$ K, which could lead to discrepancies in the calculation of \ce{^7Li} production.

We employ the method of synthesis population to estimate the Galactic Li yields \citep{2024ApJ...969..160L,2024RAA....24j5007H,2020ApJ...890...69L,2023RAA....23b5021Z}. Based on Chomiuk \& Povich (\citeyear{2011AJ....142..197C}), the star formation rate in the Milky Way is 1.9$M_\odot$yr$^{-1}$. Assuming the initial mass function (IMF) of Kroupa (\citeyear{2001MNRAS.322..231K}), we create $10^7$ single stars, in which 1.6\% have masses between 3.5 and 7.5$M_\odot$. By the linear interpolation in the models with $Z$=0.014 in Tab.\ref{tab:1}, we can estimate the Li yields ($Y_{\rm{Li}}$) produced by these intermediate-mass AGB stars. Assuming a constant star formation rate and a 13.7 Gyr age for the Milky Way, we can estimate total Li yields produced by these AGB stars in the Galactic ISM is about 15$M_\odot$. In our non-IGW models, the total Li yield produced by these AGB stars is 8$M_{\odot}$, which is half of IGW models. Comparison to the estimates of Venture (\citeyear{2000A&A...363..605V}) and Romano (\citeyear{2001A&A...374..646R}), the total Li yield in our IGW-model is higher than 10 times theirs, which is given in Tab.\ref{tab:2}. Hernanz (\citeyear{1996ApJ...465L..27H}) and Molaro (\citeyear{2016MNRAS.463L.117M}) estimated that there are about 150$M_\odot$ Li in the ISM of the Milky Way. The contribution of Li produced by our $3.5 \sim 7.5$$M_\odot$ AGB stars to the total Li in the ISM is about 10\%. Although it may be smaller than the contribution of Li produced by nova \citep{2024ApJ...971....4G}, its contribution is also non-negligible.
\begin{table}[H]
    \centering
    \caption{Yields(${M}_\odot$) and Contribution(\%) of various sources to Li in the Galactic ISM.}
    \begin{tabular}{@{}lll@{}}
        \toprule
        Sources & Yields of Li & contribution rate \\ \midrule
        AGB (our model) & 15 M$_\odot$ & 10\% \\
        Nova(Gao 2024) & 100 M$_\odot$ & 70\% \\
        Nova(Ruckya 2017) & 17 M$_\odot$ & 10\% \\
        AGB (Romano 2001) & 1 M$_\odot$ & 0.7\% \\
        AGB (Venture 2001) & 1.5 M$_\odot$ & 1\% \\
        \bottomrule
        \label{tab:1}
    \end{tabular}
\end{table}
\section{Conclusions}
We demonstrate that IGW excited by the He-shell ignition events in AGB stars can induce extra-mixing in the radiative zone between the convective thermal pulse and the convective envelope. Using MESA involving an extra-mixing triggered by IGW, we investigate the Li-rich AGB stars. In intermediate-mass stars where the HBB mechanism can occur, the IGW mixing extends the mixing region below the convective envelope can create an extra transport channel for \ce{^7Be}. This allows more internal \ce{^7Be} in the radiative zone between the convective thermal pulse and the convective envelope to be transported to the convective envelope. The positive effects of IGW mixing diminish with increasing initial mass. For higher-mass stars, the temperature at the bottom of the convective envelope is sufficiently high to sustain \ce{^7Be} production and facilitate its transport to the surface. Consequently, IGW mixing exerts a weaker positive effect on Li yields compared to lower-mass stars. Using IMF and a stellar birth rate in the Milky Way, the total Li yield from AGB stars is calculated to be 15$M_\odot$, approximately twice that of the non-IGW case. In previous models, AGB stars were not considered significant contributors of Li to the Galactic under normal mass-loss rates. Significant contributions of Li to the galaxy from AGB stars require artificially adjusted mass-loss rates, which are often inconsistent with observational constraints. In our model, under normal mass-loss rates, AGB stars make a significant contribution to Li in the Galactic. Although still lower than the 70\% contribution from nova eruptions. But for the ISM, this is a contribution that cannot be ignored, reaching 10\%.

\begin{acknowledgements}
This work received the support of the Nation Natural Science Foundation of China under grants 12163005, U20311204, 12373038, and 12288102; the Natural Science Foundation of Xinjiang No. 2022TSYCLJ0006 and 2022D01D85; the Outstanding Graduate Innovation Project of Xinjiang University with No.XJDX2025YJS039; and the science research grants from the china Manned Space Project with No. CMS-CSST-2021-A10.
\end{acknowledgements}

\bibpunct{(}{)}{;}{a}{}{,}
\bibliographystyle{aa}
\bibliography{aa}

\begin{thebibliography}{92}
\expandafter\ifx\csname natexlab\endcsname\relax\def\natexlab#1{#1}\fi

\bibitem[{{Alastuey} \& {Jancovici}(1978)}]{1978ApJ...226.1034A}
{Alastuey}, A. \& {Jancovici}, B. 1978, \apj, 226, 1034

\bibitem[{{Asplund} {et~al.}(2009){Asplund}, {Grevesse}, {Sauval}, \&
  {Scott}}]{2009ARA&A..47..481A}
{Asplund}, M., {Grevesse}, N., {Sauval}, A.~J., \& {Scott}, P. 2009, \araa, 47,
  481

\bibitem[{{Bloecker}(1995)}]{1995A&A...297..727B}
{Bloecker}, T. 1995, \aap, 297, 727

\bibitem[{{Burgers}(1969)}]{1969fecg.book.....B}
{Burgers}, J.~M. 1969, {Flow Equations for Composite Gases}

\bibitem[{{Cameron}(1955)}]{1955ApJ...121..144C}
{Cameron}, A.~G.~W. 1955, \apj, 121, 144

\bibitem[{{Cameron} \& {Fowler}(1971)}]{1971ApJ...164..111C}
{Cameron}, A.~G.~W. \& {Fowler}, W.~A. 1971, \apj, 164, 111

\bibitem[{{Canuto} {et~al.}(1996){Canuto}, {Goldman}, \&
  {Mazzitelli}}]{1996ApJ...473..550C}
{Canuto}, V.~M., {Goldman}, I., \& {Mazzitelli}, I. 1996, \apj, 473, 550

\bibitem[{{Canuto} \& {Mazzitelli}(1991)}]{1991ApJ...370..295C}
{Canuto}, V.~M. \& {Mazzitelli}, I. 1991, \apj, 370, 295

\bibitem[{{Charbonnel} \& {Talon}(2005)}]{2005Sci...309.2189C}
{Charbonnel}, C. \& {Talon}, S. 2005, Science, 309, 2189

\bibitem[{{Chomiuk} \& {Povich}(2011)}]{2011AJ....142..197C}
{Chomiuk}, L. \& {Povich}, M.~S. 2011, \aj, 142, 197

\bibitem[{{Choplin} {et~al.}(2024){Choplin}, {Siess}, {Goriely}, \&
  {Martinet}}]{2024Galax..12...66C}
{Choplin}, A., {Siess}, L., {Goriely}, S., \& {Martinet}, S. 2024, Galaxies,
  12, 66

\bibitem[{{Coc} {et~al.}(2014){Coc}, {Uzan}, \&
  {Vangioni}}]{2014JCAP...10..050C}
{Coc}, A., {Uzan}, J.-P., \& {Vangioni}, E. 2014, \jcap, 2014, 050

\bibitem[{{Cyburt} {et~al.}(2010){Cyburt}, {Amthor}, {Ferguson}, {Meisel},
  {Smith}, {Warren}, {Heger}, {Hoffman}, {Rauscher}, {Sakharuk}, {Schatz},
  {Thielemann}, \& {Wiescher}}]{2010ApJS..189..240C}
{Cyburt}, R.~H., {Amthor}, A.~M., {Ferguson}, R., {et~al.} 2010, \apjs, 189,
  240

\bibitem[{{Deepak} \& {Reddy}(2019)}]{2019MNRAS.484.2000D}
{Deepak} \& {Reddy}, B.~E. 2019, \mnras, 484, 2000

\bibitem[{{Ferguson} {et~al.}(2005){Ferguson}, {Alexander}, {Allard}, {Barman},
  {Bodnarik}, {Hauschildt}, {Heffner-Wong}, \& {Tamanai}}]{2005ApJ...623..585F}
{Ferguson}, J.~W., {Alexander}, D.~R., {Allard}, F., {et~al.} 2005, \apj, 623,
  585

\bibitem[{{Fowler} {et~al.}(1955){Fowler}, {Burbidge}, \&
  {Burbidge}}]{1955ApJ...122..271F}
{Fowler}, W.~A., {Burbidge}, G.~R., \& {Burbidge}, E.~M. 1955, \apj, 122, 271

\bibitem[{{Freytag} {et~al.}(1996){Freytag}, {Ludwig}, \&
  {Steffen}}]{1996A&A...313..497F}
{Freytag}, B., {Ludwig}, H.~G., \& {Steffen}, M. 1996, \aap, 313, 497

\bibitem[{{Fulbright}(2000)}]{2000AJ....120.1841F}
{Fulbright}, J.~P. 2000, \aj, 120, 1841

\bibitem[{{Gao} {et~al.}(2024){Gao}, {Zhu}, {L{\"u}}, {Yu}, {Li}, {Liu}, \&
  {Guo}}]{2024ApJ...971....4G}
{Gao}, J., {Zhu}, C., {L{\"u}}, G., {et~al.} 2024, \apj, 971, 4

\bibitem[{{Gao} {et~al.}(2022){Gao}, {Zhu}, {Yu}, {Liu}, {Lu}, {Shi}, \&
  {L{\"u}}}]{2022A&A...668A.126G}
{Gao}, J., {Zhu}, C., {Yu}, J., {et~al.} 2022, \aap, 668, A126

\bibitem[{{Garaud} \& {Kulenthirarajah}(2016)}]{2016ApJ...821...49G}
{Garaud}, P. \& {Kulenthirarajah}, L. 2016, \apj, 821, 49

\bibitem[{{Garc{\'\i}a-Hern{\'a}ndez}
  {et~al.}(2007){Garc{\'\i}a-Hern{\'a}ndez}, {Garc{\'\i}a-Lario}, {Plez},
  {Manchado}, {D'Antona}, {Lub}, \& {Habing}}]{2007A&A...462..711G}
{Garc{\'\i}a-Hern{\'a}ndez}, D.~A., {Garc{\'\i}a-Lario}, P., {Plez}, B.,
  {et~al.} 2007, \aap, 462, 711

\bibitem[{{Garc{\'\i}a-Hern{\'a}ndez}
  {et~al.}(2013){Garc{\'\i}a-Hern{\'a}ndez}, {Zamora}, {Yag{\"u}e},
  {Uttenthaler}, {Karakas}, {Lugaro}, {Ventura}, \&
  {Lambert}}]{2013A&A...555L...3G}
{Garc{\'\i}a-Hern{\'a}ndez}, D.~A., {Zamora}, O., {Yag{\"u}e}, A., {et~al.}
  2013, \aap, 555, L3

\bibitem[{{Garcia Lopez} \& {Spruit}(1991)}]{1991ApJ...377..268G}
{Garcia Lopez}, R.~J. \& {Spruit}, H.~C. 1991, \apj, 377, 268

\bibitem[{{He} {et~al.}(2024){He}, {Zhu}, {L{\"u}}, {Li}, {Guo}, {Liu}, \&
  {Gao}}]{2024RAA....24j5007H}
{He}, G., {Zhu}, C., {L{\"u}}, G., {et~al.} 2024, Research in Astronomy and
  Astrophysics, 24, 105007

\bibitem[{{Hernanz} {et~al.}(1996){Hernanz}, {Jose}, {Coc}, \&
  {Isern}}]{1996ApJ...465L..27H}
{Hernanz}, M., {Jose}, J., {Coc}, A., \& {Isern}, J. 1996, \apjl, 465, L27

\bibitem[{{Herwig}(2000)}]{2000A&A...360..952H}
{Herwig}, F. 2000, \aap, 360, 952

\bibitem[{{Herwig} {et~al.}(1999){Herwig}, {Bl{\"o}cker}, {Langer}, \&
  {Driebe}}]{1999A&A...349L...5H}
{Herwig}, F., {Bl{\"o}cker}, T., {Langer}, N., \& {Driebe}, T. 1999, \aap, 349,
  L5

\bibitem[{{Herwig} {et~al.}(2023){Herwig}, {Woodward}, {Mao}, {Thompson},
  {Denissenkov}, {Lau}, {Blouin}, {Andrassy}, \& {Paul}}]{2023MNRAS.525.1601H}
{Herwig}, F., {Woodward}, P.~R., {Mao}, H., {et~al.} 2023, \mnras, 525, 1601

\bibitem[{{Iben}(1967)}]{1967ApJ...147..624I}
{Iben}, Icko, J. 1967, \apj, 147, 624

\bibitem[{{Iglesias} \& {Rogers}(1993)}]{1993ApJ...412..752I}
{Iglesias}, C.~A. \& {Rogers}, F.~J. 1993, \apj, 412, 752

\bibitem[{{Iglesias} \& {Rogers}(1996)}]{1996ApJ...464..943I}
{Iglesias}, C.~A. \& {Rogers}, F.~J. 1996, \apj, 464, 943

\bibitem[{{Itoh} {et~al.}(1979){Itoh}, {Totsuji}, {Ichimaru}, \&
  {Dewitt}}]{1979ApJ...234.1079I}
{Itoh}, N., {Totsuji}, H., {Ichimaru}, S., \& {Dewitt}, H.~E. 1979, \apj, 234,
  1079

\bibitem[{{Iwamoto} {et~al.}(2004){Iwamoto}, {Kajino}, {Mathews}, {Fujimoto},
  \& {Aoki}}]{2004ApJ...602..377I}
{Iwamoto}, N., {Kajino}, T., {Mathews}, G.~J., {Fujimoto}, M.~Y., \& {Aoki}, W.
  2004, \apj, 602, 377

\bibitem[{{Karakas} \& {Lattanzio}(2014)}]{2014PASA...31...30K}
{Karakas}, A.~I. \& {Lattanzio}, J.~C. 2014, \pasa, 31, e030

\bibitem[{{Karakas} \& {Lugaro}(2016)}]{2016ApJ...825...26K}
{Karakas}, A.~I. \& {Lugaro}, M. 2016, \apj, 825, 26

\bibitem[{{Kroupa}(2001)}]{2001MNRAS.322..231K}
{Kroupa}, P. 2001, \mnras, 322, 231

\bibitem[{{Kumar} \& {Reddy}(2020)}]{2020JApA...41...49K}
{Kumar}, Y.~B. \& {Reddy}, B.~E. 2020, Journal of Astrophysics and Astronomy,
  41, 49

\bibitem[{{Lagarde} \& {Charbonnel}(2010)}]{2010sf2a.conf..253L}
{Lagarde}, N. \& {Charbonnel}, C. 2010, in SF2A-2010: Proceedings of the Annual
  meeting of the French Society of Astronomy and Astrophysics, ed.
  S.~{Boissier}, M.~{Heydari-Malayeri}, R.~{Samadi}, \& D.~{Valls-Gabaud}, 253

\bibitem[{{Lau} {et~al.}(2012){Lau}, {Doherty}, {Gil-Pons}, \&
  {Lattanzio}}]{2012MSAIS..22..247L}
{Lau}, H. H.~B., {Doherty}, C.~L., {Gil-Pons}, P., \& {Lattanzio}, J.~C. 2012,
  Memorie della Societa Astronomica Italiana Supplementi, 22, 247

\bibitem[{{Lecoanet} \& {Quataert}(2013)}]{2013MNRAS.430.2363L}
{Lecoanet}, D. \& {Quataert}, E. 2013, \mnras, 430, 2363

\bibitem[{{Li} {et~al.}(2024){Li}, {Zhu}, {L{\"u}}, {Li}, {Liu}, {Guo}, {Yu},
  \& {Lu}}]{2024ApJ...969..160L}
{Li}, Z., {Zhu}, C., {L{\"u}}, G., {et~al.} 2024, \apj, 969, 160

\bibitem[{{Lodders} {et~al.}(2009){Lodders}, {Palme}, \&
  {Gail}}]{2009LanB...4B..712L}
{Lodders}, K., {Palme}, H., \& {Gail}, H.~P. 2009, Landolt B\&ouml;rnstein, 4B,
  712

\bibitem[{{L{\"u}} {et~al.}(2020){L{\"u}}, {Zhu}, {Wang}, {Liu}, {Li}, {Xie},
  \& {Liu}}]{2020ApJ...890...69L}
{L{\"u}}, G., {Zhu}, C., {Wang}, Z., {et~al.} 2020, \apj, 890, 69

\bibitem[{{Lu} {et~al.}(2025){Lu}, {Zhu}, {L{\"u}}, {Guo}, {Li}, \&
  {Zhao}}]{2025PhRvD.111j3004L}
{Lu}, X., {Zhu}, C., {L{\"u}}, G., {et~al.} 2025, \prd, 111, 103004

\bibitem[{{Martell} {et~al.}(2021){Martell}, {Simpson}, {Balasubramaniam},
  {Buder}, {Sharma}, {Hon}, {Stello}, {Ting}, {Asplund}, {Bland-Hawthorn}, {De
  Silva}, {Freeman}, {Hayden}, {Kos}, {Lewis}, {Lind}, {Zucker}, {Zwitter},
  {Campbell}, {{\v{C}}otar}, {Horner}, {Montet}, \&
  {Wittenmyer}}]{2021MNRAS.505.5340M}
{Martell}, S.~L., {Simpson}, J.~D., {Balasubramaniam}, A.~G., {et~al.} 2021,
  \mnras, 505, 5340

\bibitem[{{Mazzitelli} {et~al.}(1999){Mazzitelli}, {D'Antona}, \&
  {Ventura}}]{1999A&A...348..846M}
{Mazzitelli}, I., {D'Antona}, F., \& {Ventura}, P. 1999, \aap, 348, 846

\bibitem[{{Molaro} {et~al.}(1997){Molaro}, {Bonifacio}, \&
  {Pasquini}}]{1997MNRAS.292L...1M}
{Molaro}, P., {Bonifacio}, P., \& {Pasquini}, L. 1997, \mnras, 292, L1

\bibitem[{{Molaro} {et~al.}(2016){Molaro}, {Izzo}, {Mason}, {Bonifacio}, \&
  {Della Valle}}]{2016MNRAS.463L.117M}
{Molaro}, P., {Izzo}, L., {Mason}, E., {Bonifacio}, P., \& {Della Valle}, M.
  2016, \mnras, 463, L117

\bibitem[{{Montalban}(1994)}]{1994A&A...281..421M}
{Montalban}, J. 1994, \aap, 281, 421

\bibitem[{{Paquette} {et~al.}(1986){Paquette}, {Pelletier}, {Fontaine}, \&
  {Michaud}}]{1986ApJS...61..177P}
{Paquette}, C., {Pelletier}, C., {Fontaine}, G., \& {Michaud}, G. 1986, \apjs,
  61, 177

\bibitem[{{Paxton}(2021)}]{2021zndo...5798242P}
{Paxton}, B. 2021, {Modules for Experiments in Stellar Astrophysics (MESA)}

\bibitem[{{Paxton} {et~al.}(2011){Paxton}, {Bildsten}, {Dotter}, {Herwig},
  {Lesaffre}, \& {Timmes}}]{2011ApJS..192....3P}
{Paxton}, B., {Bildsten}, L., {Dotter}, A., {et~al.} 2011, \apjs, 192, 3

\bibitem[{{Paxton} {et~al.}(2013){Paxton}, {Cantiello}, {Arras}, {Bildsten},
  {Brown}, {Dotter}, {Mankovich}, {Montgomery}, {Stello}, {Timmes}, \&
  {Townsend}}]{2013ApJS..208....4P}
{Paxton}, B., {Cantiello}, M., {Arras}, P., {et~al.} 2013, \apjs, 208, 4

\bibitem[{{Paxton} {et~al.}(2015){Paxton}, {Marchant}, {Schwab}, {Bauer},
  {Bildsten}, {Cantiello}, {Dessart}, {Farmer}, {Hu}, {Langer}, {Townsend},
  {Townsley}, \& {Timmes}}]{2015ApJS..220...15P}
{Paxton}, B., {Marchant}, P., {Schwab}, J., {et~al.} 2015, \apjs, 220, 15

\bibitem[{{Paxton} {et~al.}(2018){Paxton}, {Schwab}, {Bauer}, {Bildsten},
  {Blinnikov}, {Duffell}, {Farmer}, {Goldberg}, {Marchant}, {Sorokina},
  {Thoul}, {Townsend}, \& {Timmes}}]{2018ApJS..234...34P}
{Paxton}, B., {Schwab}, J., {Bauer}, E.~B., {et~al.} 2018, \apjs, 234, 34

\bibitem[{{Paxton} {et~al.}(2019){Paxton}, {Smolec}, {Schwab}, {Gautschy},
  {Bildsten}, {Cantiello}, {Dotter}, {Farmer}, {Goldberg}, {Jermyn}, {Kanbur},
  {Marchant}, {Thoul}, {Townsend}, {Wolf}, {Zhang}, \&
  {Timmes}}]{2019ApJS..243...10P}
{Paxton}, B., {Smolec}, R., {Schwab}, J., {et~al.} 2019, \apjs, 243, 10

\bibitem[{{P{\'e}rez-Mesa} {et~al.}(2019){P{\'e}rez-Mesa}, {Zamora},
  {Garc{\'\i}a-Hern{\'a}ndez}, {Osorio}, {Masseron}, {Plez}, {Manchado},
  {Karakas}, \& {Lugaro}}]{2019A&A...623A.151P}
{P{\'e}rez-Mesa}, V., {Zamora}, O., {Garc{\'\i}a-Hern{\'a}ndez}, D.~A.,
  {et~al.} 2019, \aap, 623, A151

\bibitem[{{P{\'e}rez-Mesa} {et~al.}(2017){P{\'e}rez-Mesa}, {Zamora},
  {Garc{\'\i}a-Hern{\'a}ndez}, {Plez}, {Manchado}, {Karakas}, \&
  {Lugaro}}]{2017A&A...606A..20P}
{P{\'e}rez-Mesa}, V., {Zamora}, O., {Garc{\'\i}a-Hern{\'a}ndez}, D.~A.,
  {et~al.} 2017, \aap, 606, A20

\bibitem[{{Plez} {et~al.}(1993){Plez}, {Smith}, \&
  {Lambert}}]{1993ApJ...418..812P}
{Plez}, B., {Smith}, V.~V., \& {Lambert}, D.~L. 1993, \apj, 418, 812

\bibitem[{{Prantzos}(2012)}]{2012A&A...542A..67P}
{Prantzos}, N. 2012, \aap, 542, A67

\bibitem[{{Press}(1981)}]{1981ApJ...245..286P}
{Press}, W.~H. 1981, \apj, 245, 286

\bibitem[{{Reimers}(1975)}]{1975psae.book..229R}
{Reimers}, D. 1975, in Problems in stellar atmospheres and envelopes., ed.
  B.~{Baschek}, W.~H. {Kegel}, \& G.~{Traving}, 229--256

\bibitem[{{Rogers} \& {Nayfonov}(2002)}]{2002ApJ...576.1064R}
{Rogers}, F.~J. \& {Nayfonov}, A. 2002, \apj, 576, 1064

\bibitem[{{Romano} {et~al.}(2001){Romano}, {Matteucci}, {Ventura}, \&
  {D'Antona}}]{2001A&A...374..646R}
{Romano}, D., {Matteucci}, F., {Ventura}, P., \& {D'Antona}, F. 2001, \aap,
  374, 646

\bibitem[{{Rukeya} {et~al.}(2017){Rukeya}, {L{\"u}}, {Wang}, \&
  {Zhu}}]{2017PASP..129g4201R}
{Rukeya}, R., {L{\"u}}, G., {Wang}, Z., \& {Zhu}, C. 2017, \pasp, 129, 074201

\bibitem[{{Sackmann} \& {Boothroyd}(1992)}]{1992ApJ...392L..71S}
{Sackmann}, I.~J. \& {Boothroyd}, A.~I. 1992, \apjl, 392, L71

\bibitem[{{Schwab}(2020)}]{2020ApJ...901L..18S}
{Schwab}, J. 2020, \apjl, 901, L18

\bibitem[{{Simonucci} {et~al.}(2013){Simonucci}, {Taioli}, {Palmerini}, \&
  {Busso}}]{2013ApJ...764..118S}
{Simonucci}, S., {Taioli}, S., {Palmerini}, S., \& {Busso}, M. 2013, \apj, 764,
  118

\bibitem[{{Singh} {et~al.}(2019){Singh}, {Reddy}, {Bharat Kumar}, \&
  {Antia}}]{2019ApJ...878L..21S}
{Singh}, R., {Reddy}, B.~E., {Bharat Kumar}, Y., \& {Antia}, H.~M. 2019, \apjl,
  878, L21

\bibitem[{{Singh} {et~al.}(2021){Singh}, {Reddy}, {Campbell}, {Kumar}, \&
  {Vrard}}]{2021ApJ...913L...4S}
{Singh}, R., {Reddy}, B.~E., {Campbell}, S.~W., {Kumar}, Y.~B., \& {Vrard}, M.
  2021, \apjl, 913, L4

\bibitem[{{Smith} \& {Lambert}(1989)}]{1989ApJ...345L..75S}
{Smith}, V.~V. \& {Lambert}, D.~L. 1989, \apjl, 345, L75

\bibitem[{{Smith} \& {Lambert}(1990)}]{1990ApJ...361L..69S}
{Smith}, V.~V. \& {Lambert}, D.~L. 1990, \apjl, 361, L69

\bibitem[{{Spite} {et~al.}(1996){Spite}, {Francois}, {Nissen}, \&
  {Spite}}]{1996A&A...307..172S}
{Spite}, M., {Francois}, P., {Nissen}, P.~E., \& {Spite}, F. 1996, \aap, 307,
  172

\bibitem[{{Stancliffe} {et~al.}(2010){Stancliffe}, {Angelou}, \&
  {Lattanzio}}]{2010IAUS..268..405S}
{Stancliffe}, R.~J., {Angelou}, G.~C., \& {Lattanzio}, J.~C. 2010, in IAU
  Symposium, Vol. 268, Light Elements in the Universe, ed. C.~{Charbonnel},
  M.~{Tosi}, F.~{Primas}, \& C.~{Chiappini}, 405--410

\bibitem[{{Stanton} \& {Murillo}(2016)}]{2016PhRvE..93d3203S}
{Stanton}, L.~G. \& {Murillo}, M.~S. 2016, \pre, 93, 043203

\bibitem[{{Starrfield} {et~al.}(2024){Starrfield}, {Bose}, {Iliadis}, {Hix},
  {Woodward}, \& {Wagner}}]{2024ApJ...962..191S}
{Starrfield}, S., {Bose}, M., {Iliadis}, C., {et~al.} 2024, \apj, 962, 191

\bibitem[{{Talon} \& {Charbonnel}(1998)}]{1998A&A...335..959T}
{Talon}, S. \& {Charbonnel}, C. 1998, \aap, 335, 959

\bibitem[{{Thoul} {et~al.}(1994){Thoul}, {Bahcall}, \&
  {Loeb}}]{1994ApJ...421..828T}
{Thoul}, A.~A., {Bahcall}, J.~N., \& {Loeb}, A. 1994, \apj, 421, 828

\bibitem[{{Timmes} \& {Swesty}(2000)}]{2000ApJS..126..501T}
{Timmes}, F.~X. \& {Swesty}, F.~D. 2000, \apjs, 126, 501

\bibitem[{{Travaglio} {et~al.}(2001){Travaglio}, {Randich}, {Galli},
  {Lattanzio}, {Elliott}, {Forestini}, \& {Ferrini}}]{2001ApJ...559..909T}
{Travaglio}, C., {Randich}, S., {Galli}, D., {et~al.} 2001, \apj, 559, 909

\bibitem[{{Uttenthaler} {et~al.}(2011){Uttenthaler}, {van Stiphout}, {Voet},
  {van Winckel}, {van Eck}, {Jorissen}, {Kerschbaum}, {Raskin}, {Prins},
  {Pessemier}, {Waelkens}, {Fr{\'e}mat}, {Hensberge}, {Dumortier}, \&
  {Lehmann}}]{2011A&A...531A..88U}
{Uttenthaler}, S., {van Stiphout}, K., {Voet}, K., {et~al.} 2011, \aap, 531,
  A88

\bibitem[{{Ventura} \& {D'Antona}(2005)}]{2005A&A...431..279V}
{Ventura}, P. \& {D'Antona}, F. 2005, \aap, 431, 279

\bibitem[{{Ventura} \& {D'Antona}(2009)}]{2009A&A...499..835V}
{Ventura}, P. \& {D'Antona}, F. 2009, \aap, 499, 835

\bibitem[{{Ventura} \& {D'Antona}(2010)}]{2010MNRAS.402L..72V}
{Ventura}, P. \& {D'Antona}, F. 2010, \mnras, 402, L72

\bibitem[{{Ventura} {et~al.}(2000){Ventura}, {D'Antona}, \&
  {Mazzitelli}}]{2000A&A...363..605V}
{Ventura}, P., {D'Antona}, F., \& {Mazzitelli}, I. 2000, \aap, 363, 605

\bibitem[{{Vescovi} {et~al.}(2019){Vescovi}, {Piersanti}, {Cristallo}, {Busso},
  {Vissani}, {Palmerini}, {Simonucci}, \& {Taioli}}]{2019A&A...623A.126V}
{Vescovi}, D., {Piersanti}, L., {Cristallo}, S., {et~al.} 2019, \aap, 623, A126

\bibitem[{{Wood} {et~al.}(1983){Wood}, {Bessell}, \&
  {Fox}}]{1983ApJ...272...99W}
{Wood}, P.~R., {Bessell}, M.~S., \& {Fox}, M.~W. 1983, \apj, 272, 99

\bibitem[{{Yan} {et~al.}(2018){Yan}, {Shi}, {Zhou}, {Chen}, {Li}, {Zhang},
  {Bi}, {Wu}, {Li}, {Guo}, {Liu}, {Gao}, {Zhang}, {Zhou}, {Li}, \&
  {Zhao}}]{2018NatAs...2..790Y}
{Yan}, H.-L., {Shi}, J.-R., {Zhou}, Y.-T., {et~al.} 2018, Nature Astronomy, 2,
  790

\bibitem[{{Zamora} {et~al.}(2014){Zamora}, {Garc{\'\i}a-Hern{\'a}ndez}, {Plez},
  \& {Manchado}}]{2014A&A...564L...4Z}
{Zamora}, O., {Garc{\'\i}a-Hern{\'a}ndez}, D.~A., {Plez}, B., \& {Manchado}, A.
  2014, \aap, 564, L4

\bibitem[{{Zhu} {et~al.}(2023){Zhu}, {L{\"u}}, {Lu}, \&
  {He}}]{2023RAA....23b5021Z}
{Zhu}, C.-H., {L{\"u}}, G.-L., {Lu}, X.-Z., \& {He}, J. 2023, Research in
  Astronomy and Astrophysics, 23, 025021

\bibitem[{{{\.Z}yczkowski} {et~al.}(1998){{\.Z}yczkowski}, {Horodecki},
  {Sanpera}, \& {Lewenstein}}]{1998PhRvA..58..883Z}
{{\.Z}yczkowski}, K., {Horodecki}, P., {Sanpera}, A., \& {Lewenstein}, M. 1998,
  \pra, 58, 883

\end{thebibliography}
\begin{appendix}
\section{Table with data for the Li yield}
\begin{table}[ht]
    \centering
    \small
    \renewcommand{\arraystretch}{0.4}
    \caption{The Li yield grid and the effect of IGW mixing on Li yield.}
    \begin{tabular}{@{}llccc@{}}
        \toprule
    Z & Mass/$M_\odot$ & $Y_{\rm{Li}}$(IGW)/$M_\odot$ & $Y_{\rm{Li}}$(no IGW)/$M_\odot$ & $Y_{\rm{Li}}$(IGW-no IGW)/$M_\odot$ \\ \midrule
  0.014 & 3.5 & -$5.1\times 10^{-9}$ & -$6.1\times 10^{-9}$ & $1.0\times 10^{-9}$  \\ \addlinespace
  0.014 & 4.0 & $1.2\times 10^{-8}$ & -$4.7\times 10^{-11}$ & $1.2\times 10^{-8}$ \\ \addlinespace
  0.014 & 4.5 & $1.9\times 10^{-8}$ & $6.6\times 10^{-9}$ & $1.2\times 10^{-9}$ \\ \addlinespace
  0.014 & 5.0 & $5.3\times 10^{-9}$ & $1.4\times 10^{-9}$ & $3.9\times 10^{-9}$ \\ \addlinespace
  0.014 & 5.5 & $3.4\times 10^{-9}$ & $1.1\times 10^{-9}$ & $2.3\times 10^{-9}$ \\ \addlinespace
  0.014 & 6.0 & $5.1\times 10^{-9}$ & $1.6\times 10^{-9}$ & $3.5\times 10^{-9}$ \\ \addlinespace
  0.014 & 6.5 & $1.4\times 10^{-8}$ & $8.1\times 10^{-9}$ & $7.1\times 10^{-9}$ \\ \addlinespace
  0.014 & 7.0 & $4.4\times 10^{-8}$ & $3.1\times 10^{-8}$ & $1.3\times 10^{-8}$ \\ \addlinespace
  0.014 & 7.5 & $1.0\times 10^{-7}$ & $9.7\times 10^{-8}$ & $3.3\times 10^{-9}$ \\ \addlinespace
  \midrule
  0.004 & 3.5 & $1.3\times 10^{-8}$ & $5.5\times 10^{-9}$ & $8.6\times 10^{-9}$ \\ \addlinespace
  0.004 & 4.0 & $1.5\times 10^{-8}$ & $8.2\times 10^{-9}$ & $7.1\times 10^{-9}$ \\ \addlinespace
  0.004 & 4.5 & $7.2\times 10^{-9}$ & $5.3\times 10^{-9}$ & $1.9\times 10^{-9}$ \\ \addlinespace
  0.004 & 5.0 & $4.9\times 10^{-9}$ & $3.6\times 10^{-9}$ & $1.3\times 10^{-9}$ \\ \addlinespace
  0.004 & 5.5 & $8.2\times 10^{-9}$ & $6.1\times 10^{-9}$ & $2.1\times 10^{-9}$ \\ \addlinespace
  0.004 & 6.0 & $6.6\times 10^{-9}$ & $5.6\times 10^{-9}$ & $9.8\times 10^{-10}$ \\ \addlinespace
  0.004 & 6.5 & $3.9\times 10^{-8}$ & $3.8\times 10^{-8}$ & $9.2\times 10^{-10}$ \\ \addlinespace
  0.004 & 7.0 & $1.5\times 10^{-7}$ & $1.5\times 10^{-7}$ & $3.2\times 10^{-9}$ \\ \addlinespace
  0.004 & 7.5 & $2.8\times 10^{-10}$& $7.1\times 10^{-10}$ & -$4.3\times 10^{-10}$ \\ \addlinespace
  \midrule
  0.0014 & 3.5 & $2.2\times 10^{-8}$ & $1.5\times 10^{-8}$ & $6.9\times 10^{-9}$ \\ \addlinespace
  0.0014 & 4.0 & $9.4\times 10^{-9}$ & $6.5\times 10^{-9}$ & $2.9\times 10^{-9}$ \\ \addlinespace
  0.0014 & 4.5 & $4.7\times 10^{-9}$ & $3.4\times 10^{-9}$ & $1.3\times 10^{-9}$ \\ \addlinespace
  0.0014 & 5.0 & $2.9\times 10^{-9}$ & $2.2\times 10^{-9}$ & $7.1\times 10^{-10}$ \\ \addlinespace
  0.0014 & 5.5 & $2.5\times 10^{-9}$ & $2.2\times 10^{-9}$ & $2.9\times 10^{-10}$ \\ \addlinespace
  0.0014 & 6.0 & $4.9\times 10^{-9}$ & $4.6\times 10^{-9}$ & $3.2\times 10^{-10}$ \\ \addlinespace
  0.0014 & 6.5 & $6.2\times 10^{-8}$ & $5.8\times 10^{-8}$ & $3.9\times 10^{-9}$ \\ \addlinespace
  0.0014 & 7.0 & $3.9\times 10^{-10}$& $5.9\times 10^{-10}$ & -$2.0\times 10^{-10}$ \\ \addlinespace
  0.0014 & 7.5 &-$6.1\times 10^{-10}$& $3.9\times 10^{-11}$ & -$6.5\times 10^{-10}$ \\ \addlinespace
  \midrule
  0.00014 & 3.5 & $1.8\times 10^{-8}$ & $1.2\times 10^{-8}$ & $6.1\times 10^{-9}$ \\ \addlinespace
  0.00014 & 4.0 & $9.6\times 10^{-9}$ & $3.2\times 10^{-9}$ & $6.4\times 10^{-9}$ \\ \addlinespace
  0.00014 & 4.5 & $5.7\times 10^{-9}$ & $2.3\times 10^{-9}$ & $3.4\times 10^{-9}$  \\ \addlinespace
  0.00014 & 5.0 & $8.4\times 10^{-10}$& $6.1\times 10^{-10}$ & $2.3\times 10^{-10}$ \\ \addlinespace
  0.00014 & 5.5 &-$8.4\times 10^{-10}$&-$6.1\times 10^{-10}$ & -$2.3\times 10^{-10}$ \\ \addlinespace
  0.00014 & 6.0 & $2.7\times 10^{-9}$ & $7.4\times 10^{-9}$ & -$4.7\times 10^{-9}$ \\ \addlinespace
  0.00014 & 6.5 & $5.7\times 10^{-8}$ & $7.2\times 10^{-8}$ & -$1.5\times 10^{-8}$ \\ \addlinespace
  0.00014 & 7.0 & -$1.3\times 10^{-9}$& $2.3\times 10^{-10}$ & -$1.1\times 10^{-9}$ \\ \addlinespace
  0.00014 & 7.5 & $3.2\times 10^{-8}$ & $1.2\times 10^{-7}$ & -$8.7\times 10^{-8}$ \\ \addlinespace
        \bottomrule
        \label{tab:2}
    \end{tabular}
  \tablefoot{Li yields $Y_{\rm{Li}}$ (${M}_\odot$) produced by the intermediate-mass AGB stars with different metallicities (the first column) and initial masses (the second column) for the models with and without IGW mixing. Li yields represent the net contribution of Li to the ISM after subtracting the initial Li mass inherited from the ISM. The third column represent Li yield with IGW mixing, which the forth column represent Li yield without IGW mixing. The last column corresponds to the third column minus the forth column.}
\end{table}
\end{appendix}
\end{document}